\documentclass{aa}  

\usepackage{graphicx}
\usepackage{txfonts}
\usepackage{url}

\usepackage{subfig}

\usepackage{natbib,twoopt}
\usepackage[breaklinks=true]{hyperref} 
\usepackage{pdflscape}
\bibpunct{(}{)}{;}{a}{}{,} 
\makeatletter
\newcommandtwoopt{\citeads}[3][][]{\href{http://adsabs.harvard.edu/abs/#3}%
{\def\hyper@linkstart##1##2{}%
\let\hyper@linkend\@empty\citealp[#1][#2]{#3}}}
\newcommandtwoopt{\citepads}[3][][]{\href{http://adsabs.harvard.edu/abs/#3}%
{\def\hyper@linkstart##1##2{}%
\let\hyper@linkend\@empty\citep[#1][#2]{#3}}}
\newcommandtwoopt{\citetads}[3][][]{\href{http://adsabs.harvard.edu/abs/#3}%
{\def\hyper@linkstart##1##2{}%
\let\hyper@linkend\@empty\citet[#1][#2]{#3}}}
\newcommandtwoopt{\citeyearads}[3][][]%
{\href{http://adsabs.harvard.edu/abs/#3}
{\def\hyper@linkstart##1##2{}%
\let\hyper@linkend\@empty\citeyear[#1][#2]{#3}}}
\makeatother

\begin{document}

   \title{Near-infrared colors of minor planets recovered from VISTA - VHS survey (MOVIS)}


   \author{M. Popescu\inst{1,2}
          \and
          J. Licandro\inst{3,4}
          \and
          D. Morate\inst{3,4}
          \and
          J. de Le\'on\inst{3,4}
          \and
          D. A. Nedelcu\inst{2,1}
          \and
          R. Rebolo\inst{3,4}
          \and
          R. G. McMahon\inst{5,6}
          \and
          E. Gonzalez-Solares\inst{5}
          \and
          M. Irwin\inst{5}
          }

   \institute{IMCCE, Observatoire de Paris, PSL Research University, CNRS, Sorbonne Universit\'es, UPMC Univ Paris 06, Univ. Lille, France
              \and
              Astronomical Institute of the Romanian Academy, 5 Cu\c{t}itul de Argint, 040557 Bucharest, Romania
              \and
              Instituto de Astrof\'{\i}sica de Canarias (IAC), C/V\'{\i}a L\'{a}ctea s/n, 38205 La Laguna, Tenerife, Spain
              \and
              Departamento de Astrof\'{\i}sica, Universidad de La Laguna, 38206 La Laguna, Tenerife, Spain
              \and
              Institute  of  Astronomy,  University  of  Cambridge,  Madingley  Road, Cambridge CB3 0HA, UK
              \and
              Kavli  Institute  for  Cosmology,  University  of  Cambridge,  Madingley, Road, Cambridge CB3 0HA, UK
             }

   \date{08 Jan 2016}

 
  \abstract
   {The Sloan Digital Sky Survey (SDSS)  and Wide-field Infrared Survey Explorer (WISE)  provide information about the surface composition of about 100\,000 minor planets. The resulting visible colors and  albedos enabled us to group them in several major classes, which are a simplified view of the diversity shown by the few existing spectra. A large set of data in the 0.8 - 2.5 $\mu m$, where wide spectral features are expected, is required to refine and  complement the global picture of these small bodies of the solar system.}
   {We aim to obtain the near-infrared colors for a large sample of solar system objects using the observations made during the  VISTA-VHS survey.}
   { We performed a serendipitous search in VISTA-VHS observations using a pipeline developed to retrieve and process the data that corresponds to solar system objects (SSo). The resulting photometric data is analyzed using color-color plots and by comparison with the known spectral properties of asteroids.} 
   { The colors and the magnitudes of the minor planets observed by the VISTA survey are compiled into three catalogs that are available online: the detections catalog (MOVIS-D), the magnitudes catalog (MOVIS-M), and the colors catalog (MOVIS-C\thanks{  The catalogs are available in electronic form at the CDS via anonymous ftp to cdsarc.u-strasbg.fr (130.79.128.5) or via \url{http://cdsweb.u-strasbg.fr/cgi-bin/qcat?J/A+A/}}). They were built using the third data release of the survey (VISTA VHS-DR3). A total of 39\,947 objects were detected, including 52 NEAs, 325 Mars Crossers, 515 Hungaria asteroids, 38\,428 main-belt asteroids, 146 Cybele asteroids, 147 Hilda asteroids, 270 Trojans, 13 comets, 12 Kuiper Belt objects and Neptune with its four satellites. The colors found for asteroids with known spectral properties reveal well-defined patterns corresponding to different mineralogies. The distributions of MOVIS-C data in color-color plots shows clusters identified with different taxonomic types. All the diagrams that use {\emph (Y-J)} color separate the spectral classes more effectively than the {\emph (J-H)} and {\emph (H-Ks)} plots used until now: even for large color errors (<0.1), the plots {\emph (Y-J)} vs {\emph (Y-Ks)} and {\emph (Y-J)} vs {\emph (J-Ks)} provide the separation between S-complex and C-complex. The end members A, D, R, and V-types occupy well-defined regions. 
   }
   {}

  \keywords{minor planets;  techniques: photometric, spectroscopic;  methods: observations, statistical}

  \maketitle
  
%

\section{Introduction}

About 700\,000 minor planets (small bodies of the solar system orbiting the Sun) are known today. They occupy a variety of orbits ranging from near-Earth t the Kuiper belt. Their study is motivated both by the fundamental science of solar system origins, and by practical reasons concerning  space exploration and the impact frequency with Earth.

The discovery of minor planets has grown almost exponentially in the past two decades thanks  to dedicated surveys. However, the physical characteristics  such as compositions \citep{2010A&A...510A..43C, 2013Icar..226..723D}, sizes \citep{2012ApJ...760L..12M, 2011ApJ...741...90M}, and masses \citep{2012P&SS...73...98C} are available only for a fraction of them. Visible colors and albedo are known for about 100\,000 asteroids thanks to SDSS \citep{2001AJ....122.2749I} and to WISE \citep{2011ApJ...731...53M}, respectively. Around 20\,000 minor planets have some spectrophotometric data in the J, H, and Ks bands from the Two-Micron Sky Survey (2MASS) \citep{2000Icar..146..161S}.  Spectroscopic data in visible wavelengths are available for $\sim$2\,500 of these objects (e.g. \citealt{2002Icar..158..106B, 2004Icar..172..179L}), while near-infrared (NIR) spectra are available for $\sim$1\,000 (e.g. \citealt{2009Icar..202..160D}).

The above mentioned datasets of visible colors and albedo only enables us to group the minor planets population in a few major classes, without reflecting the diversity revealed by the small number of spectra. Even  this sparse data shows a compositional mixing between different orbital groups, which points to a turbulent history of the solar system \citep{2014Natur.505..629D}. As pointed out by \cite{2014Natur.505..629D}, the next step to improve this knowledge is to refine the compositions for a large sample  of minor planets.

In this work we intend to address this question using the data obtained by an ongoing survey in the NIR performed by the VISTA (Visible and Infrared Survey Telescope for Astronomy) telescope \citep{2015A&A...575A..25S}. This is a 4-m class wide field survey telescope located at ESO's Cerro Paranal Observatory in Chile. Having the main mirror with a diameter of 4.1 m, it is the world's largest survey telescope. The VISTA telescope is equipped with a near infrared camera with a 1.65$\degr$ field of view, and with the broad band filters Z, Y, J, H, and Ks. Six large public surveys are running on VISTA\citep{2015A&A...575A..25S}: UltraVISTA, VIKING (VISTA Kilo-Degree Infrared Galaxy Survey), VMC (VISTA Magellanic Survey), VVV (VISTA Variables in the Via Lactea), VHS (VISTA Hemisphere Survey), and VIDEO (VISTA Deep Extragalactic Observations Survey). These surveys cover different areas of the sky to different depths to provide data for a variety of astrophysical fields, ranging from low-mass stars to large-scale structure of the Universe. 

\begin{figure}[!ht]
\begin{center}
\includegraphics[width=9cm]{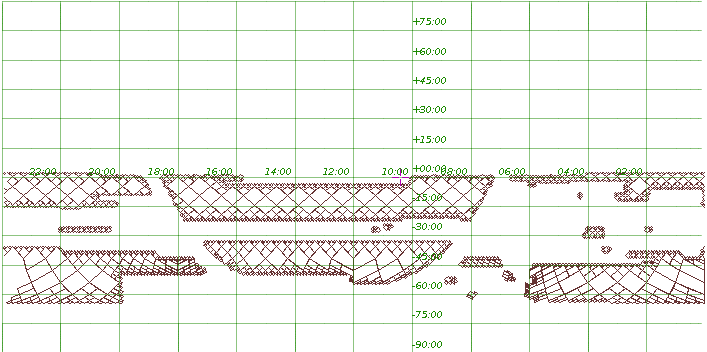}
\caption{The sky area (equatorial equidistant cylindrical projection) covered by VISTA VHS-DR3 \citep{2013Msngr.154...35M}.}
\label{areecoverd}
\end{center}
\end{figure}

The VHS survey \citep{2013Msngr.154...35M} covers the largest sky area and aims to image the entire southern hemisphere, $\sim$ 19\,000 square degrees. The resulting data is almost 4 magnitudes deeper than earlier 2MASS \citep{2006AJ....131.1163S} and DENIS \citep{1994Ap&SS.217....3E} surveys, thus we expect to find up to 100\,000 minor planets when the entire survey will be completed. In this paper we used the third data release of the survey (VISTA VHS-DR3), which imaged a field of 8\,239 square degrees, representing $\sim$20\% of the total sky area (Fig.~\ref{areecoverd}).
This corresponds to a progress of $\sim$ 40\% of the planned survey.

We compiled the VISTA-VHS sources associated with SSo in a set of catalogs called Moving Objects from VISTA survey (MOVIS). This article describes the pipeline used to obtain the spectrophotometric and astrometric data  of these SSo and discusses the results in the context of their spectral types and taxonomies. The paper is organized as follows: a general description of the VISTA-VHS observing strategy and data products is introduced in Sect. 2; the algorithms forming the MOVIS pipeline are presented in Sect. 3; the structure of the catalogs is explained in Sect. 4; and, the results are analyzed in Sect. 5.

\section{VISTA Hemisphere Survey(VHS): observations and data products}

This section briefly introduces the observation strategy and the data-flow of the VISTA-VHS survey. A detailed description is presented by \cite{2012A&A...548A.119C}. 

\subsection{Survey observations}

Infrared imaging deals with a large number of artifacts and it is strongly affected by the rapid sky variability. Therefore, imaging is commonly performed in a jitter mode: the observation of a region of the sky is broken up in short exposures and the telescope is moved slightly between them. Once reduced, these images are combined to form a single image, allowing the correction of most of these effects (bias level, bad pixels, flat-fielding and sky subtraction). 

The wide field of NIR imaging of VISTA survey is done using a mosaic camera composed of 16 large 2048 x 2048 pixels Raytheon VIRGO infrared detectors. The plate scale of a detector is 0.34\arcsec/pixel. The detectors are separated by a space of 10.4 arcmin (which represents $\sim$90\% of the detector size) in the X direction and by a space of 4.9 arcmin (which represents $\sim$42.5\% of the detector size) in the Y direction \citep{2012A&A...548A.119C}.

A single exposure reduced frame is called a {\emph normal} frame. These frames can be stacked (co-added) together with small offsets in position using a dithering pattern. This {\emph stack} frame reduce the effects of bad pixels and increase the signal to noise ratio (S/N). The area of sky covered by the pixels of a normal frame is 0.6 square degrees. 
  
\begin{figure}[!ht]
\begin{center}
\includegraphics[width=9cm]{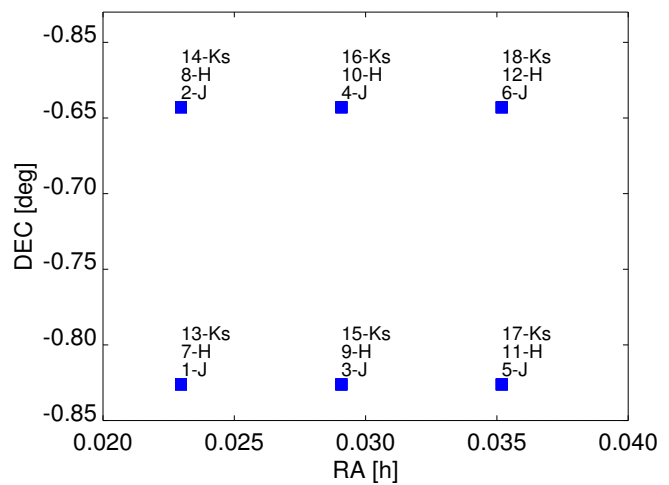}
\caption{The sequence of telescope movements to image an area of the sky using three different filters: J (1-6), H (7-12), and Ks (12-18).}
\label{TSequence}
\end{center}
\end{figure}

The VISTA basic filled survey area is a {\emph tilestack} (mosaic) image.  It is made up of a sequence of six stacks obtained by shifting the pointing of the telescope: three pointings separated by 47.5\% of a detector size are made in the Y-direction of the camera and for each of them two pointings are made, which are separated  by 95\% of a detector size in  the  X-direction .  In Fig.~\ref{TSequence} we show a typical sequence of stacks, with Y aligned with right ascension (RA) and X aligned with declination (DEC), which produces the tilestack images in J, H, and Ks.

The VISTA-VHS survey includes three programs: (1) VHS-GPS (Galactic Plane Survey), which will cover a region of $\sim$8\,200 square degrees with J and Ks filters;  (2) VHS-DAS (Dark Energy Survey), which will observe $\sim$4\,500 square degrees with J, H, and Ks filters; (3) VHS-ATLAS,which will observe $\sim$5\,000 square degrees evenly divided between North and South Galactic caps with Y, J, H, and Ks filters. Thus, an object can be observed with two up to four of these filters.
   
\begin{figure*}[!ht]
\begin{center}
\includegraphics[width=17.5cm]{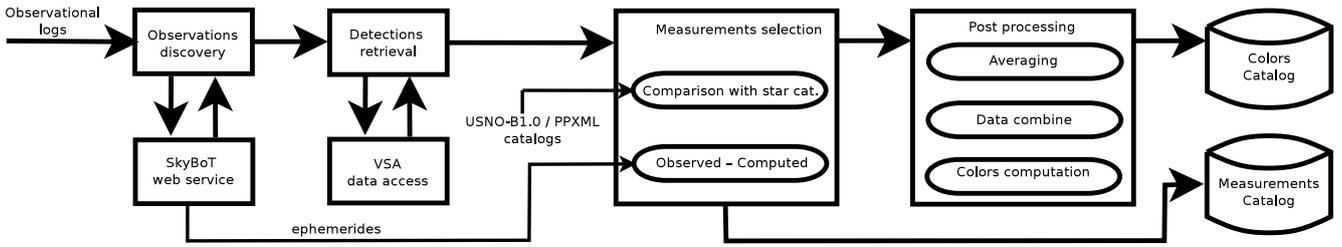}
\caption{Block diagram of MOVIS solar system objects recovering pipeline.}
\label{blockdiagram}
\end{center}
\end{figure*}

\subsection{Data products}

The observations are processed with the VISTA Data Flow System \citep{2004SPIE.5493..411I, 2010ASPC..434...91L}. These science 
products are available at the ESO Science Archive Facility and at the VISTA Science Archive \citep{2012A&A...548A.119C}.

Stack and tilestack images \citep{2004SPIE.5493..411I, 2010ASPC..434...91L} are processed to obtain the astrometric positions and the photometry of each of the sources detected in the images. This information is stored in the detection catalogs available at VISTA Science Archive (VSA), which is a component of the VISTA Data Flow System (VDFS). The VDFS is the pipeline that accomplishes the end-to-end requirements of the VISTA survey: from on-site monitoring of the quality of the data acquired, removal of instrumental artifacts, astrometric and photometric calibration, to accessibility and user-specified data products \citep{2004SPIE.5493..401E,2004SPIE.5493..423H,2004SPIE.5493..411I}.

The data products are stored in a relational database management system (RDBMS). A set of interfaces enables us to access the data \citep{2012A&A...548A.119C}. The detailed description of the tables corresponding to each data release is provided online\footnote{\url{http://horus.roe.ac.uk/vsa/www/vsa_browser.html}}. The VSA tables used in this article are {\emph Multiframe} and {\emph vhsDetection}. The information was retrieved using the Freeform SQL interface provided on the website.

The {\emph Multiframe} table contains the observing logs for each image identified by the {\emph multiframeID}. This provides the date and time of observation, the coordinates of the center of the field, the filter used, a quality grade of the image and other details related to the observing strategy and conditions. These specifications are given for all types of frames (normal, stack, and tilestack).

The {\emph vhsDetection} table contains the photometric and astrometric measurements of the objects imaged in the stack frames. The raw extraction attributes are provided for each detection. A detection is a measurement of an object obtained from a single stack frame.

The dataset corresponding to VISTA VHS-DR3 that was used in this work was obtained between November 4, 2009 and October 20, 2013. It contains 9\,276 stack frames obtained with Y filter, 32\,796 with J filter, 11\,760 with H filter, and 32\,730 with Ks filter.   

\section{Solar system objects recovering pipeline}

By covering a large area of the sky, VISTA-VHS survey imaged many SSo and included their measurements in the catalogs. To retrieve the Y, J, H, and Ks photometric data, as well as the accurate astrometry of these objects, we developed a pipeline to make the association between the SSo predicted positions and the detections found in {\emph vhsDetection} catalog. This section describes the methods used to obtain the astrometry, the photometry, and the colors of the SSo included in VISTA-VHS survey catalogs and the completeness and reliability of the results. Fig.~\ref{blockdiagram} shows the schematic of this recovering pipeline (called MOVIS pipeline).
  
The algorithm is divided into several steps: 1) find the SSo that were imaged by the survey; 2) retrieve their corresponding astrometric and photometric measurements from the {\emph vhsDetection} table; 3) validate the detections based on observed minus computed (O-C) positions, and by comparing them with USNO-B1 star catalog ; 4) post-process the information to obtain the colors and spectrophotometry of each object. The final products of the pipeline are split into three catalogs, each one addressing data combined in a different manner: the detections catalog (MOVIS-D), the spectrophotometric catalog (MOVIS-M), and the colors catalog (MOVIS-C).

\subsection{Solar system objects discovery observations in the survey fields}

Compared with the background stars, SSo appear as moving sources. Finding these objects in an observation field requires a cross-matching of the right ascension (RA) and declination (DEC) of the known SSo (computed for the accurate time-stamp of the observation), and the coordinates of the imaged field of view \citep[see][for an example of this process]{2013AN....334..718V}.

\begin{figure}[!ht]
\begin{center}
\includegraphics[width=9cm]{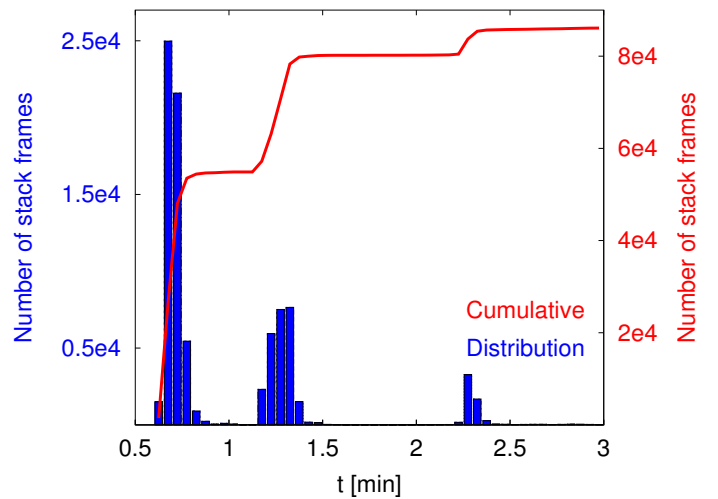}
\caption{The distribution of the time intervals in which the stack frames were obtained.}
\label{tdistrib}
\end{center}
\end{figure}

The movement rate of SSo ranges from milli arc seconds per minute, for trans-Neptunian objects, to several arc seconds per minute, for near-Earth asteroids (NEAs). The typical value for a main-belt object (MBO) is 0.3 \arcsec/min. If the movement of the object during the exposure, given by the exposure time of the image multiplied by its movement rate, is smaller than the astronomical seeing, the object appears as a point-like source. Otherwise, the object may form a trail on the image. For an image obtained by co-adding multiple exposures, which is the case for some types of frames from  the VISTA survey, the object can appear as a double  lobed object or as multiple objects if the observing time interval (the interval between the beginning of first exposure and the end of last exposure that are co-added) multiplied by movement rate is larger than the astronomical seeing. Considering this constraint, the most appropriate type of image to obtain photometry and astrometry of SSo is the stack frame. The histogram of the observing time intervals of the stack frames used in this work is shown in Fig.~\ref{tdistrib}. Taking into account the seeing and the aperture size, we can obtain accurate photometry for most of the objects with an apparent movement rate less than $\sim$2\arcsec/min, sufficient to cover the MBOs and most of the NEAs. 
  
We retrieved all the observing logs that correspond to stack frames for which the observation type was the {\emph object}, and which were not deprecated (although different degrees of deprecation may be considered for further versions of the data). For the VHS-DR3 release, this selection provided 86\,502 entries (images). The information obtained includes the coordinates (RA and DEC) of the center of each field and the accurate timing (given as MJD - modified Julian Day) of the observation for each frame identified by {\emph multiframeID} (which is also a key parameter  for other tables used by our pipeline). Based on these parameters (RA, DEC, MJD), we used Simple Cone Search (SCS) web-service provided by SkyBoT \citep{2006ASPC..351..367B}, via the VO-IMCCE website\footnote{\url{http://vo.imcce.fr/webservices/skybot/?conesearch}}. The SkyBoT cone-search method enabled us to retrieve the computed position of all the known solar system objects located in a specific field of view. For each retrieved object, it provides information regarding its designation (name, number, temporary designation), ephemeris (RA, DEC, V magnitude, phase angle, movement rates, and orbital uncertainties), and dynamical classification according to the characteristics of its orbit.
  
We automatically queried  the SkyBoT service (using SOAP protocol) for all stack frames, and we logged all the objects predicted to be in these images with V magnitude brighter than 21. This is the limiting magnitude required to obtain the photometric errors lower than 0.1 (as shown in Section 3.4), and it was estimated by considering the spectral behavior of G2V stars. The code was designed to avoid the overload of the SkyBoT server (we typically sent about 300 queries per hour). The cone search was done for a radius of 3\,000\arcsec around the center of the field to accommodate the field of view of the image and some possible orbital uncertainty. A monitoring routine detected the incomplete or empty answers and re-ran those queries again.
  
A total of 68\,237 objects with an apparent magnitude of V < 21 were predicted to be imaged by VISTA-VHS survey. A set of 62\,340 from this total number had  orbital uncertainties lower than 10\arcsec.
  
\subsection{Detections retrieval and validation}
  
To retrieve the photometry and the astrometry of the SSo predicted to be imaged, we used the enhanced version of the Freeform SQL provided by the VISTA Data Flow System for each individual stack frame. A table summarizing the information obtained from the SkyBoT was cross-matched with the {\emph vhsDetection} table using SQL commands. The cross-matching implied a squared search box centered at the predicted position. The side of the box was $6\cdot\sigma _u$ (where $\sigma _u$ is the orbital uncertainty), but no less than 2\arcsec. We considered  objects having $\sigma _u\leq 10\arcsec$ to limit the objects with apparent neighbors in the searched area, for which a separate algorithm is required. The total number of the retrieved detections  was 331\,852,  corresponding to 47\,666 objects. For each detection, all the information contained in the {\emph vhsDetection} was obtained.
  
The first step for sorting the retrieved data is the removal of the deprecated measurements. A total of 46\,880 detections were removed since they were marked as deprecated, saturated, having quality issues, or as noise. This information is contained in the following parameters of the {\emph vhsDetection} table: $deprecated\neq0$, $saturatCorr\neq0$, $ppErrBits\neq0$, respectively $class=0$.

To remove the misidentifications owing to background source confusion, a cross-matching with the USNO-B1 star catalog was made. The cross-matching was performed using the Multiple Cone Search option from TOPCAT \citep{2005ASPC..347...29T}. The search was done within a radius of 1\arcsec. A total of 11\,673 detections from the retrieved data were associated with stars and consequently removed. 
  
\begin{figure}[!ht]
\begin{center}
\includegraphics[width=9cm]{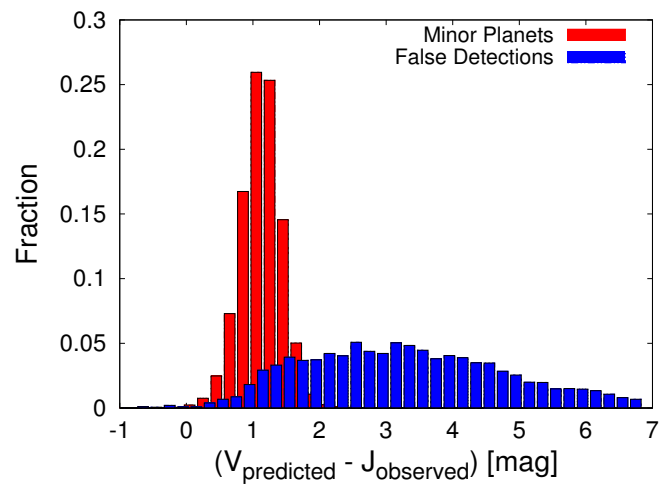}
\caption{The difference between predicted V magnitudes and observed J magnitudes for solar system objects detected in VISTA-VHS survey (88\,491 detections). The identified misassociations (the SSo which overlap with a background source) are plotted for comparison (4\,859 detections).}
\label{vpredjobs}
\end{center}
\end{figure}

The lack of a star catalog association is not a sufficient confirmation for a minor planet identification. \cite{2000Icar..146..161S} note that a way to eliminate the confusions between minor planets and background sources is to look at the difference between the predicted magnitudes and observed magnitudes. They found that identifications having V-J greater than 3 are likely detections of background sources. In Fig.~\ref{vpredjobs}  we show the distribution of the V-J for the valid detections of SSo (those detections that fulfilled all the above described criteria) compared with the distribution of V-J of those detections overlapping with an identified background source. We note that the V-J distribution of the valid detections of minor planets is centered at 1.17 with a standard deviation of 0.3, which is in agreement with \cite{2000Icar..146..161S} who noted that detections with V-J greater than 3 are likely background sources. This confirms our selection. On the other hand, the distribution of the minor planets overlapping with an identified background source is spread over a broad range of values (centered at 3.47 with a standard deviation of 1.8 ), tending to be constant between 1.5 and 4. 

On average we found five observations per night for each object. This enabled us to compute the median value of the magnitudes considering all these observations. Typically, they are spread over an interval of 20 minutes. We could assume that only one or two of these detections overlapped with a background source, thus having a very different magnitude than the rest of 3-4 detections. This assumption is justified by the fact that the average apparent movement rate is $\sim0.3$\arcsec/min which implies a position change of $\sim6$\arcsec. In this way we can remove those detections that are outside of 1.9 magnitudes (this value was chosen based on distribution from Fig.~\ref{vpredjobs}) of the median value. This value was considered sufficient to avoid overlapping with the object's intrinsic photometric variation. 
  
A strong criterion for removing the misidentifications relies on O-C. The computed coordinates are precise in the limit of the accuracy of their orbital parameters. Depending on the object, the uncertainty of the coordinates varies between several tens of milli arc seconds up to 10 arc seconds (objects with higher uncertainty were filtered out). The displacement owing to orbital uncertainty is the same (within a limit smaller than the astrometric accuracy of the observations) for all the observations of an object performed over a night, thus it can be computed as the median value of the O-C of all the observations. The detections with O-C larger than the O-C median value $\pm$0.3\arcsec are most probably misidentifications. The interval of $\pm$0.3\arcsec was selected considering the comparison between VHS positions and the VLBI radio reference frame\footnote{\url{http://www.eso.org/sci/observing/phase3/data_releases/vhs_dr2.pdf}}. This criterion which computes the alignment of the observations, applies on both RA and DEC coordinates, and it also validates the moving rate and direction of the object.
  
Another criterion for removing poor quality detections relies on the profile of the minor planet in the stack image. As explained in Section 3.1, for objects with large movement rate (NEAs), or for stack frames composed of exposures that are sparse in time, the object can appear as double lobbed or as multiple objects. By considering the average seeing of $\sim$1\arcsec and the aperture radius of 1\arcsec, we removed  detections that had their apparent movement rate multiplied by the stack time interval larger than 1.0\arcsec.
  
The number of valid measurements that remains after all the selection criteria discussed above were applied is 230\,375 ($\sim 69\%$ of the total number of detections). If a measurement fails one of these criteria, it is kept only in the detection catalog and it is flagged accordingly (see description provided in Annex~\ref{MOVISDDes}). These measurements can be used for particular purposes, in which case a different post-processing can be applied, or some degree of deprecation can be accepted. For  further statistical interpretation, we consider only the valid detections.

All the astrometric positions corresponding to valid detections were sent to the Minor Planet Center \footnote{\url{http://www.minorplanetcenter.net/}} - the worldwide location for receipt and distribution of positional measurements of minor planets. The survey received the observatory code {\emph W91-VHS-VISTA, Cerro Paranal}, and all detections were validated.

\subsection{Post-processing of the data}

Post-processing of the Y, J, H, and Ks  photometric data is required to derive compositional characterization of the observed objects. The comparison of minor planet magnitudes obtained with different filters and, in some cases, on different nights, is difficult since it needs to take into account the brightness variations that are due to object rotation and due to different heliocentric and geocentric distances. However, for statistical reasons, the following assumptions can be made:  
  \begin{itemize}
  \item { Brightness variations due to changes in heliocentric and geocentric distances can be neglected over a single night. This is true for most of the objects, except some of the NEAs.}
  \item {For statistical reasons the lightcurve variations are ignored. The error introduced can be estimated by taking into account the periods of more than 5\,500 minor planets plotted against their size, available at the Minor Planet Center\footnote{\url{http://www.minorplanetcenter.net/light_curve2/images/lcdb_all.png}}. \cite{2009Icar..202..134W} show that objects larger than 200 m have rotational periods larger than 2.4 hrs,  which is the case for most of the objects reported here. By considering the average interval in which a color was obtained as being 10 minutes, it implies a upper limit for the shift of the colors by 30\% of the lightcurve amplitude. However, we note that because the observations with all four filters are typically made  in 20 min, a similar shift owing to light curve variation will be introduced to each color, if they were obtained in roughly the same time interval.}
  \item{ Asteroids surfaces are compositionally homogeneous.}
  \item{ Phase angle effect on the colors can be neglected.}
  \end{itemize}
  
The observation strategy implies that each object is typically observed twice with each filter, although few objects may be imaged 1, 3, 4, or 6 times \citep{2012A&A...548A.119C}. If the observations are performed within 15 minutes, the averaging of the measurements obtained with the same filter will increase the signal to noise ratio.
  
\begin{figure}[!ht]
\begin{center}
\includegraphics[width=9cm]{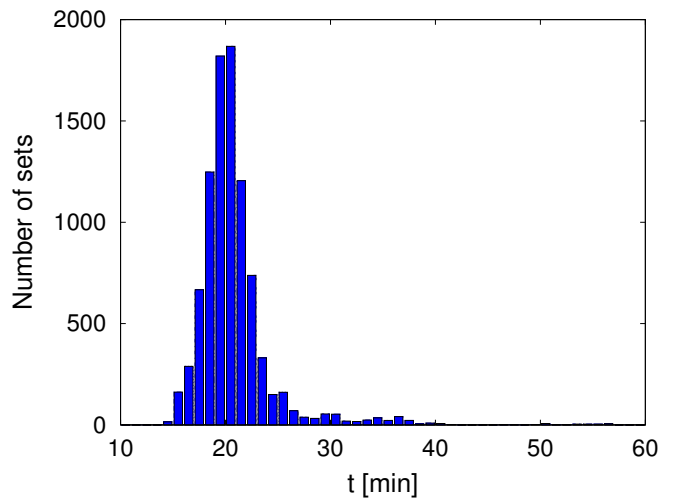}
\caption{The distribution of the time intervals in which the observations with all four filters were obtained.}
\label{magint}
\end{center}
\end{figure}

The main constraint for joining the spectrophotometric data is related to the time interval in which the observations were performed. To overcome this constraint, the algorithm was designed to join observations available in different filters with the minimum lapse of time between hem. The distribution of time intervals of the spectrophotometric sets containing all four filters is centered at $\sim$20 minutes (Fig.~\ref{magint}).
 
\begin{figure}[!ht]
\begin{center}
\includegraphics[width=9cm]{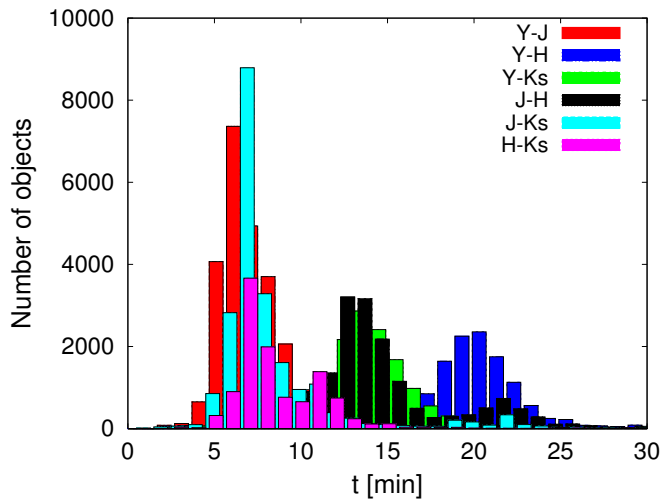}
\caption{The time interval between the two observations used to compute the colors {\emph (Y-J)}, {\emph (Y-H)}, {\emph (J-H)}, {\emph (J-Ks)}, and {\emph (H-Ks)}. The average value and standard deviation corresponding to each distribution are reported in the text.}
\label{colint}
\end{center}
\end{figure}
  
The color computation is performed by  considering observations closest in time. This minimizes the possible errors introduced by the lightcurve variation and possible sky variability. Figure~\ref{colint} shows the distribution of the time intervals between the observations used to compute different colors. These intervals depend on the observational strategy, and their mean values and standard deviations are: $t_{avg}^{Y-J}=7.6$, $\sigma_{t}^{Y-J}=2.4$;  $t_{avg}^{Y-H}=19.9$, $\sigma_{t}^{Y-H}=3.5$;  $t_{avg}^{Y-Ks}=14.1$, $\sigma_{t}^{Y-Ks}=2.7$;  $t_{avg}^{J-H}=14.6$, $\sigma_{t}^{J-H}=4.0$;  $t_{avg}^{J-Ks}=7.9$, $\sigma_{t}^{J-Ks}=3.7$;  $t_{avg}^{H-Ks}=8.2$, $\sigma_{t}^{H-Ks}=2.9$ minutes.  We filtered out those colors obtained on intervals larger than 30 minutes (which represents $t_{avg}^{Y-H} + 3\sigma_{t}^{Y-H}$ -- relative to the worst case, and includes the majority of the data). 
  
This strategy of computation implies that subtracting two colors to obtain the third one may not lead to the same result as reported in the catalog because different observations may enter in the computation of each of them. Statistically, the two results are comparable within the reported error. Moreover, about 90\% of these differences are smaller than 0.05 magnitudes.
  
The colors of minor planets obtained on different dates were merged to obtain the most complete set available for each object and averaged to improve the signal to noise ratio (in case of multiple observations of the same color). The averaging of values of the same color obtained multiple times is performed only if the  errors are comparable (within a factor of $\sqrt{2}$), otherwise the values with large errors are discarded, since these are probably affected by poor observing conditions.

\begin{figure}
\begin{center}
\includegraphics[width=9cm]{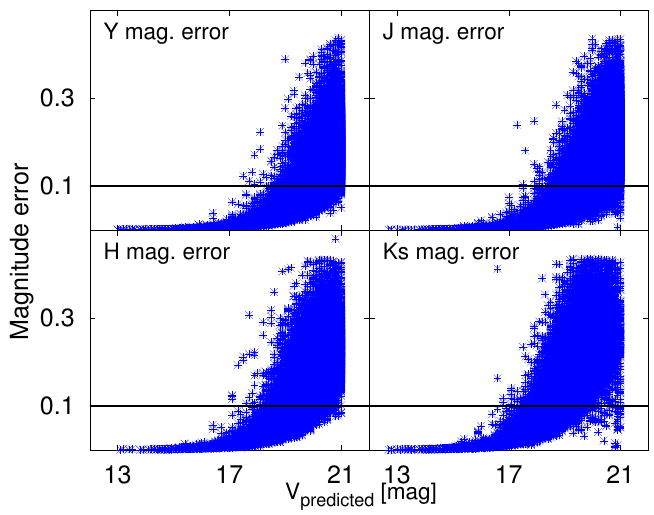}
\end{center}
\caption{Distribution of the errors associated with each magnitude relative to the predicted apparent V magnitude for MOVIS-D data. The horizontal black line corresponds to an error of 0.1 magnitudes.}
\label{snr}
\end{figure}

\begin{figure*}
\begin{center}
\includegraphics[width=18cm]{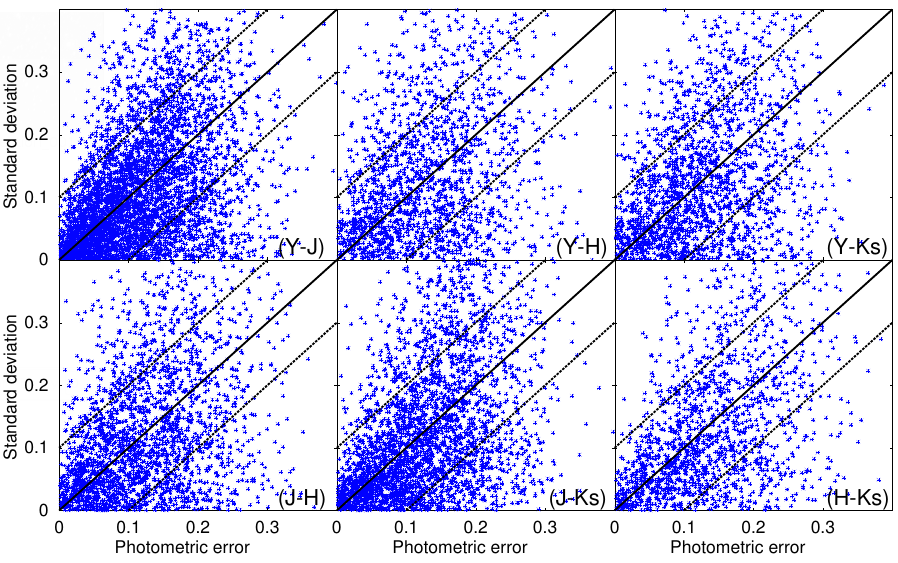}
\end{center}
\caption{Comparison between the spread of values of the color of an object observed on multiple nights (measured by standard deviation) and the averaged photometric errors of these values.}
\label{ErrorComparison}
\end{figure*}

\subsection{Completeness and reliability}
  
The completeness of the dataset can be inferred by considering the following arguments: 1) the sky area covered by the data release of the survey as shown by Fig.~\ref{areecoverd}; 2) the list of known SSo and their orbital uncertainty (the SkyBoT web service used to prepare this article, worked with the dynamical properties of asteroids issued from the 10/2014 version of the ASTORB database); 3) the detection limit of the VISTA photometric pipeline; and 4) the capability of the MOVIS pipeline to associate the detections with the corresponding SSo.
  
To overcome the constraint introduced by the orbital uncertainty, the prediction of SSo positions was made on a circular area with a radius of R = 3\,000\arcsec, centered over each stack frame. This is $\sim$ 30\arcsec \ larger than the diagonal of the field of view of the camera.
 
The photometric precision is determined by the source brightness, the photometric quality of the observing night, the detector efficiency, and the exposure time. Depending on the VHS sub-survey, the exposure time varied between 5 and 15 seconds. The magnitude limit of the survey can be inferred by plotting the photometric errors versus the predicted magnitude for the detections marked as valid by the pipeline (Fig.~\ref{snr}). The magnitude limit depends on the filter: the photometric data with magnitude errors smaller than 0.1 can be obtained with Y, J, and H filters for V magnitudes brighter than 21, and in the Ks band for V magnitudes brighter than 20. Considering these constraints and the fact that we aimed to obtain data for photometric assessments and statistical analysis, the cutoff for the predicted V magnitude was 21 (justified by the data shown in Fig.~\ref{snr}). We cannot exclude the fact that there are objects with magnitudes fainter than our cutoff (V=21), which can be detected  (either because of their predicted magnitude uncertainty, or that can be detected just in Y or J bands). These objects will be investigated later in an updated catalog version, since they can provide valuable astrometric information.
  
The fact that we considered only those objects with an orbital uncertainty lower than 10\arcsec removes 5\,897 out of 68\,237 predicted objects. A separate algorithm is under development to find objects with an uncertainty larger than 10\arcsec.

Overall, 47\,666 objects were found in the {\emph vhsDetection}. The number of objects found is lower than  predicted because a predicted position may not have an associated detection. This can be explained by taking into account the gaps between the detectors. The area considered for predicting the objects is 2.18 square degrees (circular area with a radius of 3000\arcsec) compared with the area covered by all the detectors, 16$\cdot$(2048 $\cdot$ 0.34/3600)$^2$ = 0.6 square degrees (16 square detectors of 2\,048 pixels with 0.34\arcsec/pixel). The number of predicted detections ($\sim$ 1\,500\,000) multiplied by the ratio of the two area suggests that about $\sim$ 410\,000 detections should be found, compared with the $\sim$331\,852 retrieved detections. The difference can be explained  by the limiting magnitude of the filters (which can be seen by the saturation around 0.4 error magnitude in the Ks filter - Fig.~\ref{snr}).
 
Before the final computation of the colors and of the spectrophotometric data, the pipeline removes those detections which are marked as deprecated (46\,880 detections), or have astrometric inconsistencies (9\,051 detections), or present aspect issues such as long trails or double lobbed, (3\,388 detections), could be misidentified, owing to close by apparent neighbor objects (37\,192 detections), or there is a single detections for an object (4\,966). After this operation, a total number of 230\,375 valid detections remains.

The reliability assessment is made using two methods: 1) comparison of the observations of the same object from multiple nights and 2) comparison with similar surveys performed in the same spectral region.
A direct comparison of the magnitudes of the same SSo observed on multiple nights cannot be done because their brightness varies significantly. However, in the hypothesis that their surfaces are compositionally homogeneous, their colors should not change (considering that the phase angle effects are negligible). Therefore, internal comparison of colors was possible by identifying 6\,941 objects observed on two separate nights, 1\,411 observed on three separate nights, 293 objects observed on four separate nights, and 85 objects observed on more than four separate nights. 

\begin{equation}
\sigma_{nights} = \sqrt{\frac{1}{N-1}\sum_{i=1}^{N} (c_i - \bar{c})^2}
\label{std}
.\end{equation}
In Eq.~\ref{std}, $N$ is the number of nights on which the object was observed, $c_i$ is the color obtained on the night $i$, and $\bar{c}$  is the averaged value of the obtained colors. This standard deviation quantifies the spread of values obtained on multiple nights for the color of an object.  

The $\sigma_{nights}$ was compared with the average value of the reported photometric error ($\bar{e_c}$) of these colors (Fig.~\ref{ErrorComparison}): $\approx68\%$ of the objects with $\bar{e_c}<0.1$ have $\sigma_{nights}< 0.1$, and $\approx23\%$ of them have $0.1<\sigma_{nights}< 0.2$. This result complies with the definition of the standard deviation. The comparison shows that the determined colors are consistent with the photometric error: the distribution of multiple observations is limited to the photometric accuracy. This is less valid for smaller photometric errors ($\leq\sim 0.05$), which can be explained by the lightcurve related effects.

Accurate comparison between the colors obtained by different surveys or by other particular observations is difficult since each survey tends to use its own set of filters (with different characteristics), and different observing and data reduction strategies. Up till now, the NIR colors of the largest number of SSo have been obtained by \cite{2000Icar..146..161S} using 2MASS survey. Their data are available in the {\emph 2MASS Asteroid and Comet catalogues} via the Planetary Data System node\footnote{\url{https://pds.nasa.gov/}}. 

\begin{equation}
\begin{split}
\begin{aligned}
&Y_V = J_2 + 0.610*(J-H)_2 \\ 
&J_V=J_2-0.077(J-H)_2\\
&H_V=H_2+0.032(J-H)_2\\
&Ks_V=Ks_2+0.010(J-Ks)_2\\
\end{aligned}
\end{split}
\label{V2MASS}
.\end{equation}
In Eq.~\ref{V2MASS} the  subscripts $V$ and $2$ indicate VISTA and 2MASS surveys, respectively. 

\begin{figure*}[!ht]
\begin{center}
\includegraphics[width=18cm,height=4.5cm]{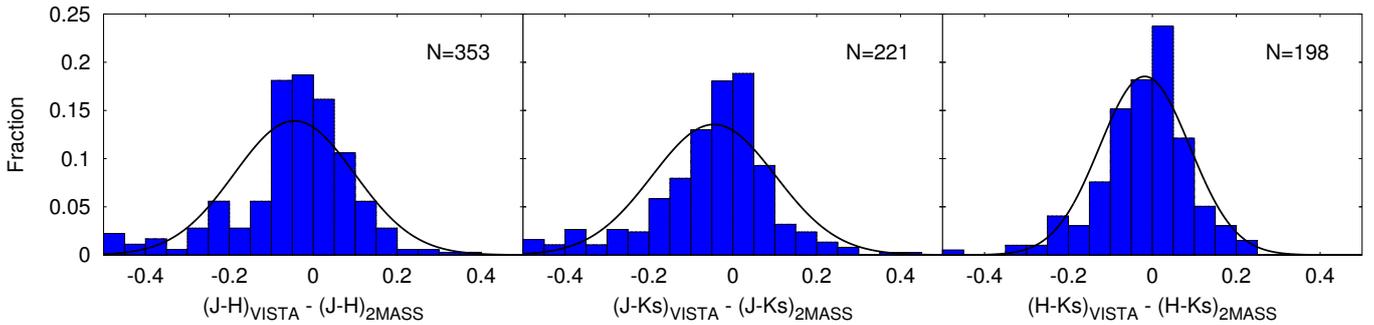}
\end{center}
\caption{Comparison between the colors obtained by VISTA and 2MASS. The number of objects is shown as a label. The histogram of differences is compared with a normal distribution.}
\label{2masscomparison}
\end{figure*}

To compare the colors of minor planets observed with J, H, and Ks filters by both 2MASS and VISTA, we need to take into account the characteristics of the filters. The expressions that relate the two filter sets were derived from a compilation of data measured on the two surveys\footnote{\url{http://casu.ast.cam.ac.uk/surveys-projects/vista/technical/photometric-properties}} Eq.~\ref{V2MASS}.
To compare the {\emph (J-H)}, {\emph (J-Ks)}, and {\emph (H-Ks)} colors of SSo observed by these two surveys, we select the measurements with photometric error less than 0.1. We found 353 SSo with {\emph (J-H)} colors, 221 with {\emph (J-Ks)} colors, and 198 with {\emph (H-Ks)} colors obtained by both surveys. We applied the expressions shown in Eq.~\ref{V2MASS} to the selected 2MASS data. The distributions of the differences between the two surveys are shown in Fig.~\ref{2masscomparison}. They correspond to the following average values ($\mu_{V2M}$) and standard deviations ($\sigma_{V2M}$): $\mu_{V2M}^{J-H} = -0.045$, $\sigma_{V2M}^{J-H} = 0.143$; $\mu_{V2M}^{J-Ks} = -0.045$, $\sigma_{V2M}^{J-Ks} = 0.147$; $\mu_{V2M}^{H-Ks} = -0.018$, $\sigma_{V2M}^{H-Ks} = 0.108$. There are several outliers -- mostly for the {\emph J-H} and {\emph (J-Ks)} colors -- which can be explained as possible misidentifications. We note that the comparison between the minor planets observed in DENIS survey and the ones obtained by 2MASS shows a similar dispersion \citep{2001A&A...375..275B, 2004A&A...423..381B}.

\begin{figure}[!ht]
\begin{center}
\includegraphics[width=9cm]{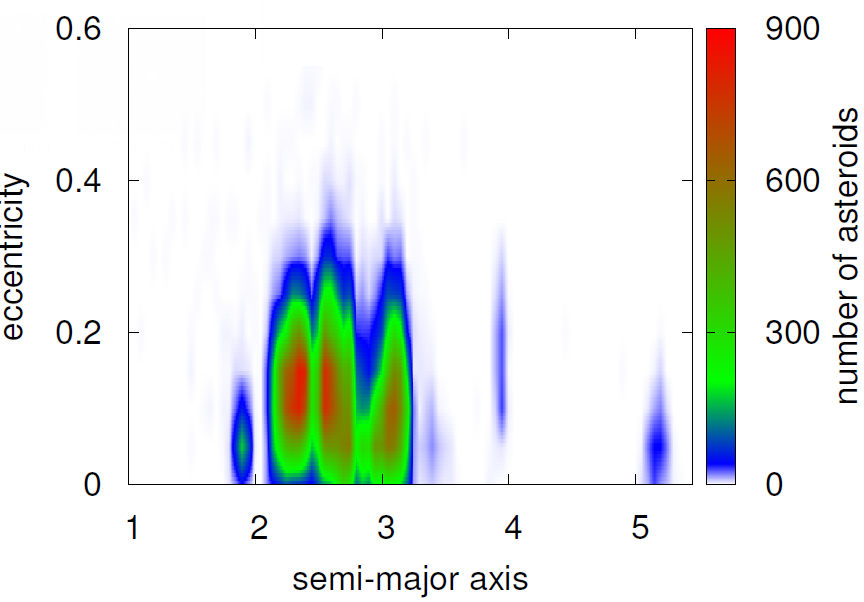}
\caption{The distribution of objects with valid measurements in the semi-major axis (given in AU) vs eccentricity plot. The color bar represents the number of asteroids in an (a,e) with a bin of size (0.05 AU, 0.05). The x axis was limited to 5.5 AU.}
\label{aeplot}
\end{center}
\end{figure}
     
\begin{figure}[!ht]
\begin{center}
\includegraphics[width=9cm]{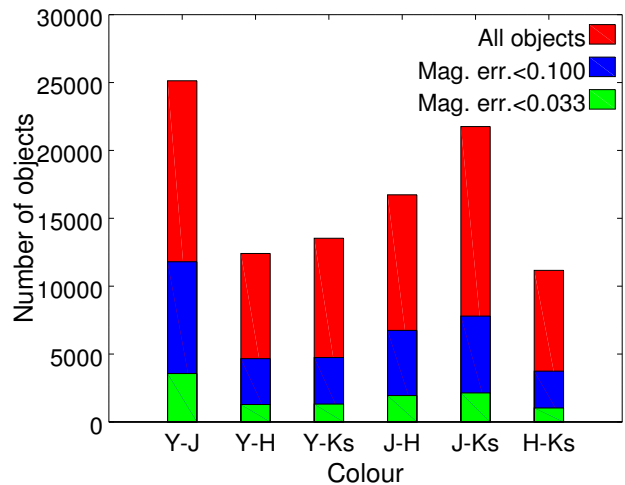}
\caption{The number of objects for which a given color was obtained.}
\label{resultvssnr}
\end{center}
\end{figure}

\section{Minor planets VISTA catalogs - MOVIS}

This section describes the final data product of the pipeline: the MOVIS catalogs. The results of the recovering pipeline are grouped in three catalogs: the detections catalog (MOVIS-D), the magnitudes catalog (MOVIS-M), and the colors catalog (MOVIS-C). The split of the data was performed with the purpose of organizing it in an efficient way for analysis. The detection information, the spectrophotometric data, and the resulting colors require different approaches to combine the data. The description of the information contained on each column is given in Table~\ref{MOVISDDes}, Table~\ref{MOVISFDes}, and Table~\ref{MOVISCDes}, for the MOVIS-D, MOVIS-M, and MOVIS-C catalogs, respectively.

MOVIS-D contains all the detections retrieved from the {\emph vhsDetection} table on the basis of their computed positions. It contains 331\,852 lines (a line provides all the information for a detection), out of which 230\,375 are valid detections,  marked with 0 in the first column of the catalog ({\emph Flag} column). The rest of the 101\,477 detections have different photometric quality issues and are marked by a non-zero flag. In  cases where a certain degree of deprecation can be accepted, these detections can provide  useful information for some particular objects.

The valid detections correspond to 39\,947 objects including 52 NEAs, 325 Mars Crossers, 515 Hungaria asteroids, 38\,428  main-belt asteroids, 146 Cybele asteroids, 147 Hilda asteroids, 270 Trojans, 13 comets, 12 Kuiper Belt objects, and Neptune with its four satellites.  The distribution of the detected objects in semi-major axis vs eccentricity plot is shown in Fig.~\ref{aeplot}. The objects with valid measurements in at least two different bands were used to build the magnitudes and the colors catalogs. The objects observed  with only a single filter are logged just in MOVIS-D file. The information provided in MOVIS-D catalog (i.e. frame ID, RA, DEC, MJD) allows to retrieve the VISTA images of any SSo using the web application form provided by VISTA science team or the ESO archive query form.

The MOVIS-M and MOVIS-C catalogs use the values of the corrected magnitudes measured with 1\arcsec aperture radius (denoted {\emph aperMag3} in {\emph vhsDetection} database). 

The magnitudes catalog (MOVIS-M) contains the measurements of an object obtained with different filters. The data is selected with the constraint to have the minimum time interval between the observations of an object in different wavelength bands. An entry contains a single result for each filter. If an object was observed on multiple nights, a separate entry is  provided for each night (called set). This catalog provides measurements for 43\,241 sets. The accuracy of the magnitudes is described by both the photometric error and the seeing. The time interval between the observations gives an indication of the effect introduced by lightcurve variations.

The colors catalog (MOVIS-C) contains the {\emph (Y-J)}, {\emph (Y-H)}, {\emph (Y-Ks)}, {\emph (J-H)}, {\emph (J-Ks)}, and {\emph (H-Ks)} colors for the 34\,998 objects found in VISTA-VHS. If an object was observed on multiple nights, the colors were merged together, as described in Section 3.3. The color errors and the time interval between the two observations used for the color computation are also provided. The number of colors for each object varies owing to different sub-survey strategies and different limiting magnitudes of the filters. Fig.~\ref{resultvssnr} shows the number of objects for which a given color was obtained.

\section{Data analysis}

 In this section, we analyze the colors of minor planets from MOVIS-C catalog by means of color-color plots. The aim is to derive information about the surface composition based on NIR colors. To achieve this, we compare our results with the known spectral properties of minor planets.

\subsection{General overview}

\begin{figure}[!ht]
\begin{center}
\includegraphics[width=9cm]{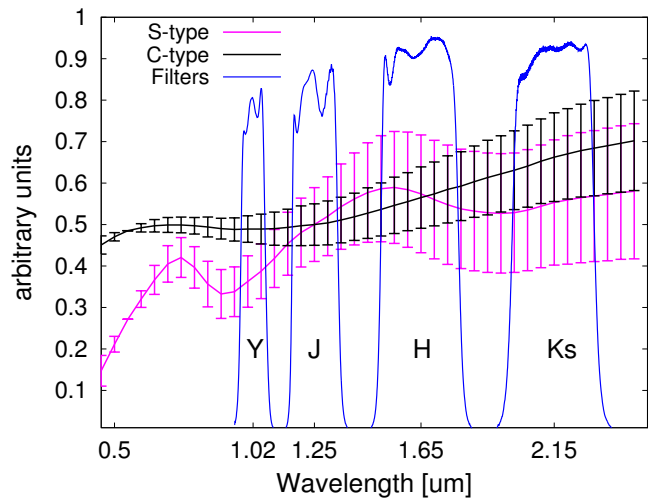}
\caption{The wavelength position of Y, J, H, and Ks filters compared with the standard S and C taxonomic types from \cite{2009Icar..202..160D}. The two reflectance spectral types (i.e. reflectance spectra are obtained as the ratio of the observed spectral data to a solar analog star spectrum) were normalized to unity at 1.25 $\mu$m and shifted down in reflectance by 0.5 for comparison.}
\label{yjhkposition}
\end{center}
\end{figure}

The visible to near-infrared (VNIR) spectral region is the most exploited for determining the surface composition of minor planets. There are three reasons for this: a) the atmosphere is relatively transparent at these wavelengths; b) the reflected component of the flux is maximal; and c) the mineralogy is the primary first-order determinant of the spectral properties \citep{1989aste.conf...98G}. The wide absorption features of asteroids in this wavelength interval can be characterized even using broadband filters \citep[e.g.][]{1988Icar...74..454H,1996A&A...305..984B,2000Icar..146..161S,2001AJ....122.2749I,2002A&A...389..641H,2008Icar..198..138P, 2013MNRAS.433.2075C, 2013Icar..226..723D, 2016Icar..268..340C}. The central wavelengths of Y, J, H, and Ks filters (Fig.~\ref{yjhkposition}) enables us to find gradients and turning points of the spectral data, as well to quantify the absorption bands \citep[e.g.][]{2002A&A...389..641H}, thus enabling us to have an approximate determination of the surface mineralogy.  For example, if we refer to typical spectra of S-type asteroids, showing the 1 and 2 $\mu$m absorption bands (which is the case of asteroid surfaces with an olivine-pyroxene composition), the Y filter samples the spectral region close to the first band minimum, the J and H filters are centered around the maximum in the NIR reflectance spectrum, and the Ks filter covers the wavelengths close to the second band minimum (Fig.~\ref{yjhkposition}).
     
Near-infrared J, H, and K photometry have proved to be a powerful tool for obtaining information about the surface composition of asteroids for a long time. \cite{1988Icar...74..454H} found a wide range of {\emph (J-H)} colors of M-type asteroids and a distinct separation of S, A, and D taxonomic classes. They also found that NEAs occupy a very large region in {\emph (J-H)} vs {\emph (H-K)} plot, corresponding to all taxonomic types.  \cite{1995Icar..114..186V} observed 56 asteroids members of Eos, Koronis, and Maria families using J, H, and K filters. They found that the objects belonging to a specific family have a similar surface composition.  \cite{2002A&A...389..641H} performed a statistical analysis of VNIR colors for 104 minor planets from the outer solar system identifying various group properties among the different dynamical populations. Recently, based on observations made with J and H filters for (624) Hektor and (762) Pulcova,  \cite{2014SoSyR..48..202G} suggest that these minor planets have heterogeneous surface composition.

 \begin{table*} 
\caption{Average NIR colors with known taxonomic type according to \cite{2002Icar..158..146B,2002Icar..158..106B}, observed by VISTA-VHS.}
\centering
\begin{tabular}{l |c c c| c c c| c c c| c c c| c c c}\hline\hline
&\multicolumn{3}{|c|}{Y-J}&\multicolumn{3}{|c|}{Y-Ks}&\multicolumn{3}{|c|}{J-Ks}&\multicolumn{3}{|c|}{J-H}&\multicolumn{3}{|c}{H-Ks} \\
Type & Mean & $\sigma$ & No. & Mean & $\sigma$ & No. & Mean & $\sigma$ & No. & Mean & $\sigma$ & No. & Mean & $\sigma$ & No.\\ \hline
A & 0.44 & 0.04 & 2  & 1.15 & 0.27 & 3  & 0.77 & 0.27 & 2  & 0.46 & -    & 1  & 0.06 & 0.02 & 2 \\
C & 0.25 & 0.06 & 60 & 0.65 & 0.10 & 62 & 0.40 & 0.11 & 75 & 0.28 & 0.08 & 40 & 0.11 & 0.08 & 44 \\
D & 0.38 & 0.03 & 11 & 0.98 & 0.15 & 11 & 0.61 & 0.11 & 12 & 0.40 & 0.08 & 8  & 0.19 & 0.08 & 8 \\
S & 0.39 & 0.08 & 63 & 0.81 & 0.15 & 63 & 0.43 & 0.11 & 86 & 0.37 & 0.12 & 57 & 0.04 & 0.08 & 59 \\
V & 0.64 & 0.09 & 5  & 0.72 & 0.10 & 5  & 0.08 & 0.08 & 5  & 0.22 & 0.07 & 5  & -0.07& 0.05 & 5 \\
X & 0.30 & 0.15 & 38 & 0.74 & 0.14 & 36 & 0.47 & 0.12 & 55 & 0.31 & 0.12 & 28 & 0.14 & 0.07 & 27\\
\hline
\end{tabular}
\label{AverageColor}     
\end{table*}

The largest set of NIR colors of minor planets was provided by \cite{2000Icar..146..161S} using 2MASS survey data. Their initial catalogs contain observations of 1\,054 asteroids and two comets. These results are discussed in {\emph (J-H)} vs {\emph (H-Ks)} space and are compared with the regions mapped by S, C, D, and A taxonomic classes. These regions were defined using data from \cite{1988Icar...74..454H}, \cite{1992Icar...99..485S} and \cite{1995Icar..114..186V}. They found that even there are significant regions in which the colors of different taxonomic types overlap, there are some areas in which a specific one dominates. The separation between S, C, and D types can be seen for data with SNR $>$ 30. \cite{2000Icar..146..161S} speculate that the larger dispersion of NIR colors obtained for low SNR data can be explained by an increasing compositional variation of objects with smaller sizes (which typically have fainter magnitudes).

Following the work presented in the papers outlined above, we used the color-color plots to analyze MOVIS-C data. To map the distribution of objects in those plots, we applied two methods: 1) plot the colors of MOVIS-C objects with an assigned taxonomic type by \cite{2002Icar..158..146B,2002Icar..158..106B} and \cite{2004Icar..172..179L}; 2) compare our results with the colors obtained from the template spectra defined by \cite{2009Icar..202..160D} for the different taxonomic classes.

The data obtained with the four filters allow the computation of six colors (Fig.~\ref{resultvssnr}). The number of color-color plots that can be generated with these colors is fifteen ($C_6^2=15$). To select the relevant ones, we considered the following arguments: (i) the {\emph (J-H)} vs {\emph (H-Ks)} plot enables us to discuss the results in the framework of previous publications; (ii) the {\emph (Y-J)} vs {\emph (H-Ks)} plot contains observations made with all four filters; (iii) because of the survey strategy and different limiting magnitudes of the filters, the majority of the data are obtained using the Y, J, and Ks filters (Fig.~\ref{resultvssnr}) ; (iv) the magnitude errors of the observations made with  H filters are larger compared with those obtained with Y and J filters. Thus, we selected four plots  for this discussion:  {\emph (J-H)} vs {\emph (H-Ks)};  {\emph (Y-J)} vs {\emph (H-Ks)};  {\emph (Y-J)} vs {\emph (J-Ks)}; and  {\emph (Y-J)} vs {\emph (Y-Ks)}.

The goal of our analysis is to identify groups of minor planets that have similar surface composition and to match them with the taxonomic types. A precise taxonomic classification gives an approach to a specific mineralogy for the corresponding object and is the first step for further studies of comparative planetology. Most of the objects presented in the MOVIS-C catalog are asteroids. As a consequence, for discussing the distributions of data in the color-color plots, we can refer to the Bus taxonomy \citep{2002Icar..158..146B,2002Icar..158..106B} and to  its extension into the NIR -- the DeMeo taxonomy \citep{2009Icar..202..160D}.

The first approach is to map the colors of asteroids with known spectral behavior. Currently, the number of minor pla\-nets with spectral observations is of the order of a few thousands \citep{2012A&A...544A.130P}. About 2\,500 spectra of main-belt asteroids were obtained by SMASSI \citep{1995Icar..115....1X}, SMASSII \citep{2002Icar..158..106B,2002Icar..158..106B},  and $S^3OS^2$ \citep{2004Icar..172..179L}. Their observations covered the wavelengths between $\sim$ 0.45 and 0.92 $\mu m$ and the spectra were classified using the Bus taxonomic system. We found 278 objects in MOVIS-C catalog with spectra observed by these large surveys. 
  
Table~\ref{AverageColor} shows the mean values and the statistical dispersions of MOVIS-C colors corresponding to asteroids belonging to the C-complex (B, C, Cb, Cg, Cgh, and Ch classes), the S-complex (S, Sk, Sl, Sq, and Sr classes), the X-complex (X, Xc, Xk, Xe), and to the end member classes A (including A and Sa types), D, and V, according to \cite{2002Icar..158..146B,2002Icar..158..106B} and \cite{2004Icar..172..179L}. This enables us to estimate the limiting color errors required to separate between different taxonomic types. For example, the Euclidean distance in the {\emph (Y-J)} vs {\emph (Y-Ks)} space between the median value of C-complex and S-complex is 0.21, which suggests that even some observations with color errors $\sim$ 0.1 can be classified within the two classes.
  
The second approach is to compare the distribution of MOVIS-C data in color-color space with the position of colors computed for the template spectra of the different taxonomic classes defined by \cite{2009Icar..202..160D}. They define 25 classes that cover the 0.45 to 2.45 $\mu$m interval. The reflectance values of the template spectra are defined for 41 wavelengths evenly spaced at 0.05 $\mu$m. The template curves were obtained by applying principal component analysis  to a set of 371 spectra of asteroids. The asteroids reflectance spectra are obtained as the ratio of the observed spectral data to a solar analog star spectrum determined in similar conditions.
 
\begin{figure*}[p]
\begin{center}
\includegraphics[width=18cm,height=5.25cm]{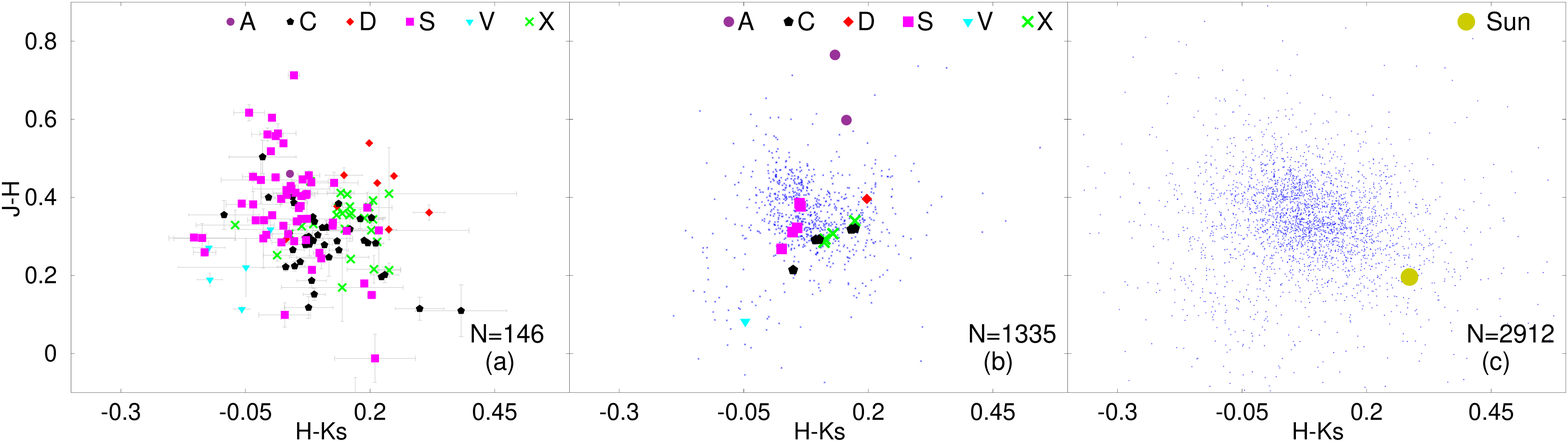}
\includegraphics[width=18cm,height=5.25cm]{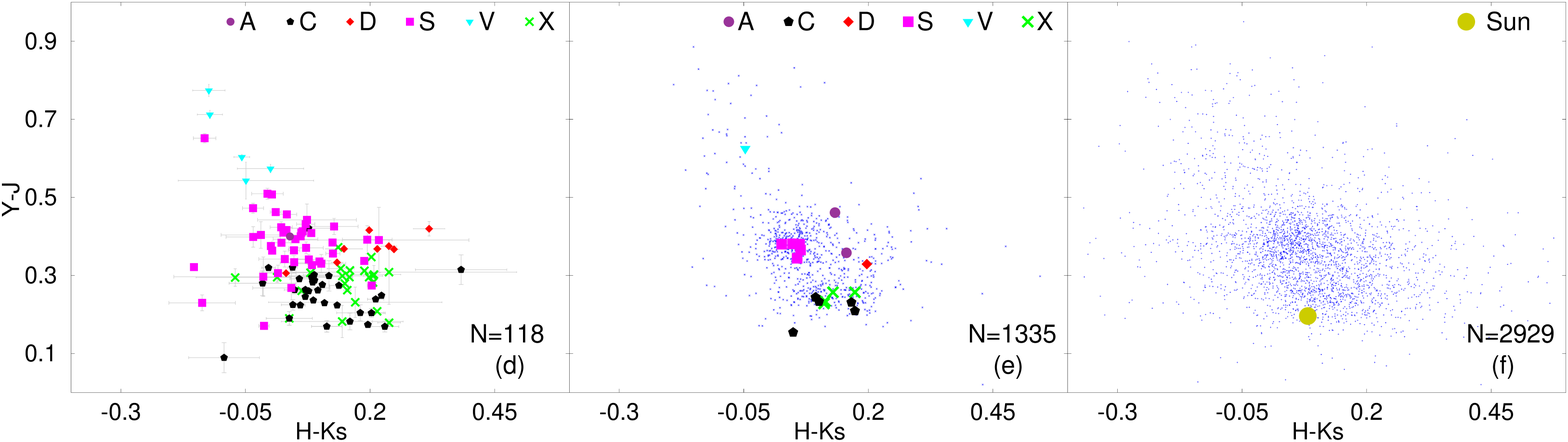}
\includegraphics[width=18cm,height=5.25cm]{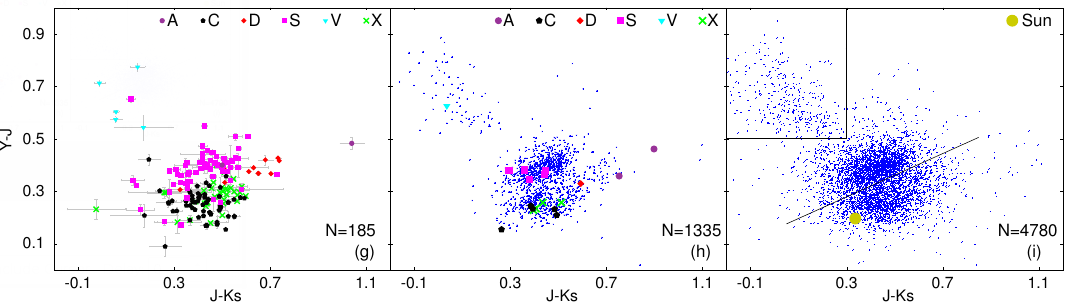}
\includegraphics[width=18cm,height=5.25cm]{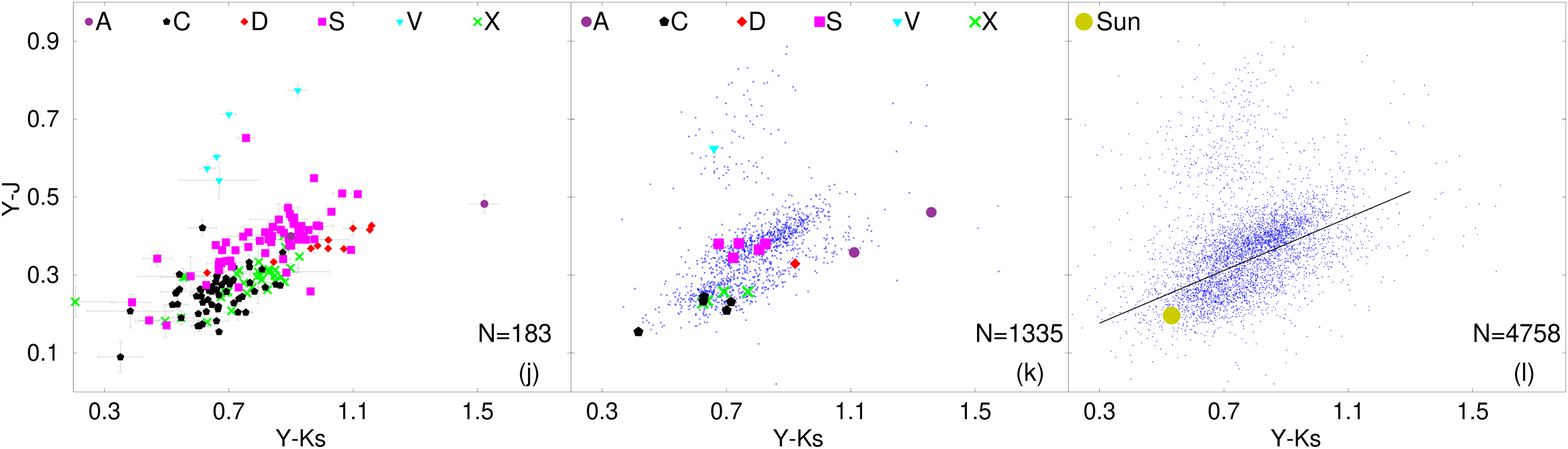}
\end{center}
\caption{Color-color plots of the MOVIS-C data: {\emph (J-H)} vs {\emph (H-Ks)} -- a,b,c; {\emph (Y-J)} vs {\emph (H-Ks)} -- d,e,f; {\emph (Y-J)} vs {\emph (J-Ks)} -- g,h,i; {\emph (Y-J)} vs {\emph (Y-Ks)} -- j,k,l. Left column: The colors of asteroids with visible spectra, having an assigned taxonomic type. Central column: the colors computed for the template spectra of the taxonomic classes from \cite{2009Icar..202..160D} compared with the MOVIS-C data with color errors less than 0.033 (C-complex includes B, C, Cb, Cg, Cgh, and Ch sub-classes and is denoted as C; S-complex includes Q, S, Sq, Sr, and Sv sub-classes and is denoted as S; X-complex includes X, Xc, Xe, and Xk sub-classes and is denoted as X; A class includes A and Sa types). Right column: the MOVIS-C data obtained with a color error less than 0.1 compared with the colors of the Sun (yellow dot).}
\label{ColorColorPlots}
\end{figure*}

We computed the equivalent of these spectral templates curves into the color domain by taking into account the response curve of the filters and the solar colors as follow: a) re-sample the template spectra to a wavelength step of 1 nm using linear interpolation; b) multiply the spectra by the filter transmission functions and integrate the result to obtain the photometric values; c) subtract them to determine the colors; d) add the colors of the Sun to the results. The obtained colors of the template spectra from the DeMeo taxonomy are shown in Table~\ref{TaxonColor}. Their discussion is made with respect to the main groups: the C-complex (B, C, Cb, Cg, Cgh, and Ch classes), the S-complex (Q, S, Sq, Sr, Sv), the X-complex (X, Xc, Xe, Xk), and to the end member classes A (including A and Sa types), D and V. Figures ~\ref{ColorColorPlots}--b,~\ref{ColorColorPlots}--d,~\ref{ColorColorPlots}--h, and ~\ref{ColorColorPlots}--k display the location of these colors compared to MOVIS-C data with color errors less than 0.033. 

We used the solar colors obtained by \cite{2012ApJ...761...16C}. Their values were found using 2MASS observations: $(J-H)_2 = 0.286$, $(J-Ks)_2 = 0.362$, $(H-Ks)_2 = 0.076$. The transformation to VISTA filters system was performed using Eq.~\ref{V2MASS}. The results are: $(Y-J)_V = 0.196 $, $(Y-Ks)_V = 0.532 $,  $(J-H)_V = 0.255 $, $(J-Ks)_V = 0.336 $, $(H-Ks)_V = 0.082 $. The errors of the colors provided by \cite{2012ApJ...761...16C} are about 0.01 magnitudes (see Table 2 from the cited article). The error of the conversion coefficients between VISTA and 2MASS system is not provided. We note that the {\emph (Y-J)} value is very uncertain as the 2MASS survey does not have the Y filter, so the solar flux in Y is computed using a linear interpolation of J, H, and Ks solar values from 2MASS. This uncertainty in the colors of the Sun translates into an offset in the color computed for the template spectra.
 
The diversity of SSo colors can be quantified by the statistical mean and variance of the MOVIS-C data with color error less than 0.033: the largest variation is $\sigma^{0.033}_{(Y-Ks)} = 0.14$, where the median value is $\overline{(Y-Ks)_{0.033}}=0.79$ (computed  over 1\,315 colors), and the smallest variation corresponds to $\sigma^{0.033}_{(H-Ks)} = 0.09$, where the median value is $\overline{(H-Ks)_{0.033}}=0.08$ (computed  over 1\,1027 colors). These statistic values do not vary significantly when decreasing the accuracy of the photometry (color errors less than 0.1): $\sigma^{0.100}_{(Y-Ks)} = 0.17$, $\overline{(Y-Ks)_{0.100}}=0.77$ (computed  over 4\,742 colors) and $\sigma^{0.100}_{(H-Ks)} = 0.08$, $\overline{(H-Ks)_{0.100}}=0.13$ (computed  over 3\,741 colors). This statistical view enables us to conclude that the set of MOVIS-C colors with errors less than 0.033 is representative for the objects shown in the catalog.
\begin{figure}[!ht]
\begin{center}
\includegraphics[width=9cm]{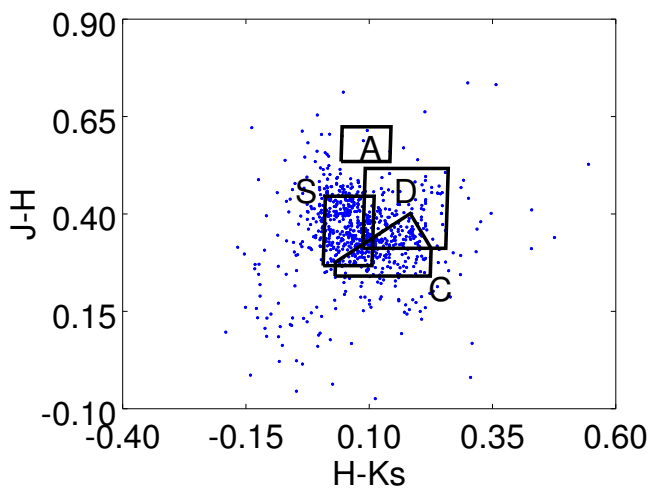} 
\end{center}
\caption{ {\emph (J-H)} vs {\emph (H-Ks)} plot: the regions found by \cite{2000Icar..146..161S} are shown with black lines, compared with MOVIS-C data with errors less than 0.033.}
\label{fig:VISTA2MASSBorders}
\end{figure}

\subsection {(J-H) vs (H-Ks)}

The {\emph (J-H)} vs {\emph (H-Ks)} plot has been used by different authors to separate the spectral classes \citep[e.g.][]{1988Icar...74..454H, 1992Icar...99..485S, 1995Icar..114..186V}. In Fig.~\ref{fig:VISTA2MASSBorders}  we present the regions mapped by \cite{2000Icar..146..161S}  in the {\emph (J-H)} vs {\emph (H-Ks)} space (corrected to the VISTA system using Eq.~\ref{V2MASS}). The MOVIS-C objects with color uncertainties smaller than 0.033 are plotted for comparison. The distribution of our data is similar to that found by \cite{2000Icar..146..161S}: most of the objects concentrate in the region defined by S, C, and D limits. The highest population density is in the region where the three classes overlap. The A class, which is the only one that is completely separate, has two objects inside its region and several others around its borders. There are also many objects nearby the outside borders of the defined S and C regions, suggesting that these regions, in our case, are larger than those mapped by \cite{2000Icar..146..161S}. Several tens of objects are located in the lower left corner of the plot, i.e. low {\emph (J-H)} and low {\emph (H-Ks)}, with a widely spread distribution. There are no points identified by \cite{2000Icar..146..161S} in this region. 

Given the small color error (<0.033) of the minor planets plotted in Fig.~\ref{fig:VISTA2MASSBorders}, the distribution in the {\emph (J-H)} vs {\emph (H-Ks)} plot can only be a consequence of a real compositional variation of these bodies. This hypothesis is also supported by the plot of MOVIS-C colors of objects with known visible spectra (Fig.~\ref{ColorColorPlots}--a). The V-types (objects with spectra similar to that of the asteroid Vesta) have low {\emph (J-H)} and {\emph (H-Ks)} colors, explaining the data located at the lower left corner of the plot (light blue). Figure~\ref{ColorColorPlots}--a also confirms the in-fill between the S and C groups. The X-types, introduced by Bus taxonomy for featureless spectra with slightly to moderately red slopes, share almost the same region with C-types. Part of them are concentrated in the limits between the C and D regions. The identified D-types appear as a distinct group, having two S-type intruders that can be explained by their large errors. 
 
The colors of the template spectra from the DeMeo taxonomy confirm the mapping of the distribution in {\emph (J-H)} vs {\emph (H-Ks)} plot (Fig.~\ref{ColorColorPlots}--b): the S, C, and X complexes are located in the dense regions, and D- and V-types are in the middle of the corner groups.
 
The partial overlapping between S, C, and D groups in {\emph (J-H)} vs {\emph (H-Ks)} plot makes it difficult to assign a taxonomic class based on these colors. Moreover, the small distance between the colors of the template spectra suggests that only data with accurate photometry (i.e. color errors less than about 0.05) can be useful for deriving information about the surface composition of the objects.  However, the {\emph (J-H)} vs {\emph (H-Ks)} is useful to identify asteroids belonging to end member classes A and V, which are separated enough from the main distribution.
 
The matching between the regions mapped by \cite{2000Icar..146..161S} and MOVIS-C data provides an additional argument for the reliability of the MOVIS pipeline.

\subsection {(Y-J) vs (H-Ks)}

The Y-filter samples the 1 $\mu$m absorption band, which is characteristic of the spectra of olivine-pyroxene compositions (Fig.~\ref{yjhkposition}). These type of objects have a steep spectral slope in the wavelength region covered by Y and J filters, which is quantified by a large value of the {\emph (Y-J)} color. Thus, it can be used to separate between the spectra having the 1 $\mu$m absorption band and featureless spectra.
  
The {\emph (Y-J)} vs {\emph (H-Ks)} plot of asteroids with know spectral classification shows the separation of the S- and C-complexes in a better way than the {\emph (J-H)} vs {\emph (H-Ks)} plot (Fig.~\ref{ColorColorPlots}--c). Objects classified as S and C/X define two clusters with almost no overlapping region. In any case, there are several interlopers which show unusual colors. Their data can be explained either as measurement artifacts, or they may have particular spectral properties. 

The asteroids classified as V-types are concentrated on the upper left side of the plot, with {\emph (Y-J)} $>$ 0.55 (Fig.~\ref{ColorColorPlots}--c). The interloper in this region (pink dot) is the asteroid numbered with 5051. This object was classified as Sr type by \cite{2002Icar..158..146B}, based on its visible spectrum. Using the curve matching methods \citep{2012A&A...544A.130P}, we found that this spectrum is matched by R-type (which has, like the V-types, a deep 1 $\mu$m band). We  also note that, according to its dynamical parameters ($a = 2.29$ AU, $e = 0.099$, $i = 6.62^\circ$), 5051 belongs to the Vesta family.

The division between the representative groups of asteroids, the S- and C/X- complexes, is shown in the {\emph (Y-J)} vs {\emph (H-Ks)} plot of colors with errors less than 0.033 (Fig.~\ref{ColorColorPlots}--d). The colors of taxonomic templates are located in the regions with high density number of objects. The data corresponding to the S-complex template spectra appear as a compact group. The C-complex, although overlapping with the X-complex, is divided into three groups, which are identified mostly in the {\emph (H-Ks)} color: B-types have {\emph (H-Ks)} = 0.05, Cgh/Ch-types have {\emph (H-Ks)} = 0.10, and C/Cb-types have {\emph (H-Ks)} = 0.17 (Table~\ref{TaxonColor}). 

The {\emph (Y-J)} vs {\emph (H-Ks)} plot of objects with color errors less than 0.10 is consistent with the distributions discussed above. The clusters corresponding to S- and C/X complexes are identifiable even for large errors  (Fig.~\ref{ColorColorPlots}--f).      

\subsection {(Y-J) vs (J-Ks)}

The reasons for introducing the {\emph (J-Ks)} color are that the expected variation for this color owing to possible spectral shapes is larger than that of the {\emph (H-Ks)} color (see Table~\ref{AverageColor}), that most of the objects in MOVIS-C catalog have observations in J and Ks bands (Fig.~\ref{resultvssnr}), and that the limiting magnitude in J band is fainter than in H band (Fig.~\ref{snr}).
 
Compared with {\emph (Y-J)} vs  {\emph (H-Ks)} diagram, the {\emph (Y-J)} vs {\emph (J-Ks)} plot of objects with known spectral classification shows a better separation between the S-complex and C-complex (Fig.~\ref{ColorColorPlots}--g). Thus, we performed a linear fit to the colors of the objects (asteroids 2952, 4733, 2374, 3170, 1734, 26879, 3885, 3675, and 4044) located just on each side of this S / C division. Their classification was obtained based on the visible spectra from the SMASS and the $S^3OS^2$ surveys .

\begin{equation}
(Y-J) = 0.412^{\pm0.046}\cdot(J-Ks) + 0.155^{\pm0.016}
\label{ymjvsjmkseq}
.\end{equation}
The linear fit obtained follows the expression - Eq.~\ref{ymjvsjmkseq}.

Regarding the separation defined by Eq.~\ref{ymjvsjmkseq}, between the S- and the C-complex, we note that there are three objects that spectrally belong to the S-complex, but located in the C-region according to their {\emph (Y-J)} and {\emph (J-Ks)} colors: 3040, 5610, and 4733. Also the asteroid 2106 is spectrally assigned to the C-complex, but it is well inside the S-complex region of the plot. By looking at the VISTA images of these objects, we found that the asteroids 3040 and 5610 overlap with a faint star in the J-filter, and these measurements are not removed by our pipeline. For the asteroids 4733 and 2106, we do not identify any artifact related to the images, which suggests either an atypical spectra in the NIR region or some particular issues that were not removed by the pipeline, such as varying atmospheric conditions or rapid lightcurve variations.
 
The plot of objects with color error less than 0.033 (Fig~\ref{ColorColorPlots}--h) shows the gap between the S-complex and the C/X-complex. In the same plot, the colors of the taxonomic templates are in agreement with this separation.
 
The asteroids with spectral properties compatible with the V-types are separated from the other spectral classes by a clear zone, defining a region with colors {\emph (Y-J)} $\geq 0.5$ and {\emph (J-Ks)} $\leq 0.3$ (Fig~\ref{ColorColorPlots}--h). Considering that the uncertainties of the colors are $<$0.033, the large spread of this group reflects the surface composition variation. Figure~\ref{ColorColorPlots}--i shows that the V-type candidates can be well identified even for large color errors (less than 0.1).
 
The {\emph (Y-J)} vs. {\emph (J-Ks)} plot also enables us to identify A-types: these are characterized by {\emph (J-Ks)} $\geq \sim 0.85$ and moderate {\emph (Y-J)}. Unfortunately there are only two objects classified as A-type with accurate MOVIS colors (Fig.~\ref{ColorColorPlots}--h).

\subsection {(Y-J) vs (Y-Ks)}

The {\emph (Y-Ks)} colors of asteroids with known visible spectral taxonomic types are spread over an interval larger than 1 magnitude (Fig.~\ref{ColorColorPlots}--j). This provides the best ratio between color variations and errors compared with other colors obtained from Y, J, H, and Ks observations.
  
\begin{equation}
(Y-J) = 0.338^{\pm0.027}\cdot(Y-Ks) + 0.075^{\pm0.02}
\label{ymjvsymkseq}
.\end{equation}

As in the case of {\emph (Y-J)} vs {\emph (J-Ks)}, the {\emph (Y-J)} vs {\emph (Y-Ks)} plot separates the primitive asteroids (C-complex) from the rocky ones (S-complex). Following the same procedure as that described for the  of Eq.~\ref{ymjvsjmkseq}, we can compute a linear fit using the colors of the asteroids located on both sides of the division line (1554, 6906, 3170, 822, 6410, 3885, 3675, 558, and 244), obtaining the expression, Eq.~\ref{ymjvsymkseq}.

With a moderate slope in the NIR region, most of the objects from the X-complex occupy the intermediate region between those belonging to the C-complex and those classified as D-types (Fig.~\ref{ColorColorPlots} --j).

The A and the V types occupy the marginal regions of the {\emph (Y-J)} vs {\emph (Y-Ks)} plot (Fig.~\ref{ColorColorPlots}--j). The only interloper in the V-type region (pink dot in Fig.~\ref{ColorColorPlots}--j) is the asteroid 5051, as discussed in Section 5.3. The D-type asteroids occupy a separate region in this color-color plot. The intruders in this region correspond to the Sa type that is characterized by a spectra with red slope in the NIR. This type can be separated from the D-types based on the presence or not of the 2 $\mu$m absorption band, detectable with the {\emph (H-Ks)} color.

\begin{figure}[!ht]
\begin{center}
\includegraphics[width=9cm,height=4.5cm]{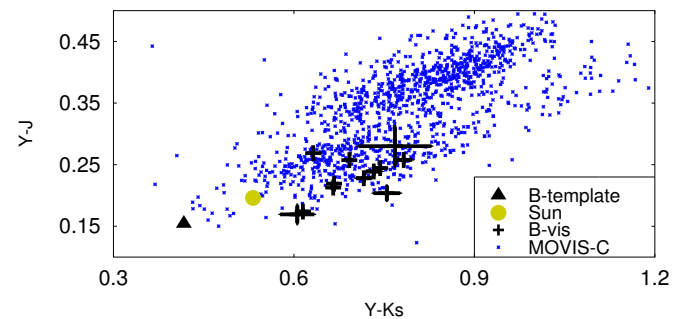}
\caption{{\emph (Y-J)} and {\emph (Y-Ks)} colors of objects classified as B-types according to their visible spectra (B-vis). The colors of the Sun and the colors of the template spectrum of B-types from \cite{2009Icar..202..160D} are shown for comparison. The MOVIS-C data with color error less than 0.033 are plotted in the background as blue dots.}
\label{Btype}
\end{center}
\end{figure}

The B-types are defined by \cite{2009Icar..202..160D} based on their negative NIR slope. As such, they should be located in the lower left corner of the {\emph (Y-J)} vs {\emph (Y-Ks)} plot. Interestingly, the colors of the B-type objects identified by SMASS and $S^3OS^2$, i.e., classified as B-types based exclusively on their visible spectra, are not located in this expected region (Fig.~\ref{Btype}). This was explained by \cite{2012Icar..218..196D} who found that asteroids classified as B-types according to their visible spectra show a continuous shape variation in their NIR spectral slopes, ranging from negative, blue slopes, to positive, moderately red slopes. Therefore, the {\emph (Y-J)} vs {\emph (Y-Ks)} plot enables us to separate the group of B-types as defined by \cite{2009Icar..202..160D}, i.e., as those having negative NIR spectral slope.

The data corresponding to the S-complex spreads almost uniformly over a very elongated region of the {\emph (Y-J)} vs {\emph (Y-Ks)} plot (Fig.~\ref{ColorColorPlots}--j), which suggests a continuous variation of the NIR slopes of the S-types. This effect can be explained by the space-weathering effects, a process that is the result of dust impacts and solar wind sputtering on the surface of atmosphereless rocky bodies, which causes a reddening of their spectral slope, a decrease in the spectral absorption band depths, and a diminishing of their albedo \citep{2001JGR...10610039H,2006Icar..184..327B,2007A&A...472..653B}.

The {\emph (Y-J)} vs {\emph (Y-Ks)} plot of objects with color error up to 0.1 is consistent with the results found for accurate observations. The computed separating line for the S- and C-complexes applies even for high errors (Fig.\ref{ColorColorPlots}--l).

\section{Conclusions and further work}

This paper describes the NIR photometric data of minor planets observed by the VISTA-VHS survey. This survey uses Y, J, H, and Ks filters for imaging the entire sky of the southern hemisphere. A total of 39\,947 SSo were detected in the VISTA VHS Data Release 3 (which covers $\sim$40\% of the planned survey sky area). The detections found include: 52 NEAs, 325 Mars Crossers, 515 Hungaria asteroids, 38\,428 main-belt asteroids, 146 Cybele asteroid, 147 Hilda asteroids, 270 Trojans, 13 comets, 12 Kuiper Belt objects, and Neptune with its four satellites.  About 34\,998 of these objects were imaged with at least two different filters.

The retrieved photometric data of SSo is provided as a set of catalogs called MOVIS. These catalogs are obtained using a pipeline that finds the objects based on their ephemeris, retrieves the data by interfacing them with VHS tables, removes the wrong associations, and does a post-processing of the data (averaging, data combination, and color computation). The correctness and reliability of the pipeline is assessed by analyzing the error distributions and by comparing the results with the 2MASS dataset.

The results are reported in three catalogs: the detections catalog (MOVIS-D), the magnitudes catalog (MOVIS-M), and the colors catalog (MOVIS-C). We presented them in Section 4 and in the annexes. These catalogs will be stored online at the CDS - Strasbourg data center.

The analysis of the near-infrared color-color plots derived from MOVIS-C data shows the large diversity among different minor planet surfaces. The patterns identified in the distribution of NIR colors correspond to different taxonomic types. The color-color plots of the asteroids with known spectral properties reveal the color intervals corresponding to various compositional types.  All the diagrams that use {\emph (Y-J)} color separate the spectral classes more accurately than has been performed until now using the {\emph (J-H)} vs {\emph (H-Ks)} plots. Even for large color uncertainties (<0.1), the plots {\emph (Y-J)} vs {\emph (Y-Ks)} and {\emph (Y-J)} vs {\emph (J-Ks)} clearly separate the asteroids belonging to the main spectroscopic  S- and C-complex and enable us to identify the end taxonomical type members A, D, R, and V types. Few outliers are outside these patterns, thus confirming the correctness of the MOVIS pipeline. 

Future work includes:
\begin{enumerate}
      \item {Continue to obtain the NIR spectrophotometric data as the survey progresses. We expect to obtain the colors of another 50\,000 objects.}
      \item {Perform a statistical analysis based on clustering methods, such as principal component analysis to provide additional information about the surface composition. Correlation of NIR colors with albedo  and visible colors (SDSS) can bring new insights about the spectral behavior over this interval. Comparison with laboratory data (i.e. with meteorites' reflectance spectra) may provide new ways to obtain the mineralogy of the observed objects.}
      \item {Obtain the distributions of objects with various dynamical parameters to provide a new mapping of the composition of the solar system.}
      \item {Update the pipeline to obtain the astrometric and the photometric information for objects with  uncertainties in their position larger than 10 arcsec.}
      \item {Provide an algorithm to taxonomically classify the objects based on their colors. This algorithm needs to take into account the distance between the classes in the color-color plots, their spread and the photometric accuracy of the objects.}
\end{enumerate}

\begin{acknowledgements}
      The article is based on observations acquired with Visible and Infrared Survey Telescope for Astronomy (VISTA). The observations were obtained as part of the VISTA Hemisphere Survey, ESO Program, 179.A-2010 (PI: McMahon).
      
      The data has been numerically analyzed with TOPCAT(\url{http://www.starlink.ac.uk/topcat/}) and GNU Octave. 
      
      We thank to Jerome Berthier for his support on using SkyBoT service.
      
      We thank to Gareth Williams and Jos\'{e} Luis Galache for their support in submitting data to Minor Planet Center and for the validation of the astrometric measurements.
      
      We thank  Francesca DeMeo and Bobby Bus for providing us the spectra used by \citep{2009Icar..202..160D} to define the taxonomy.
      
      The work of M. Popescu was supported by a grant of from Campus Atlantico Tricontinental (Programa Talento Tricontinental 2014 - \url{http://www.ceicanarias.com/}) and by a grant of the Romanian National Authority for  Scientific  Research --  UEFISCDI,  project number PN-II-RU-TE-2014-4-2199.
      
      J. de Le\'on acknowledges support from the Instituto de Astrof\'isica de Canarias.
      
      D. Morate acknowledges the Spanish MINECO for the financial support in the form of a Severo-Ochoa Ph.D. fellowship.
      
      J. Licandro, D. Morate, J. de Le\'on, and M. Popescu acknowledge support from the project ESP2013-47816-C4-2-P (MINECO, Spanish Ministry of Economy and Competitiveness). 
      
      The work of D. A. Nedelcu was supported by a grant of the Romanian National Authority for  Scientific  Research -- UEFISCDI,  project number PN-II-RU-TE-2014-4-2199.
      
      We thank  Beno\^{i}t Carry for the constructive and helpful suggestions.
      
\end{acknowledgements}

\bibliographystyle{aa}
\bibliography{MPVISTA.bib}

\newpage
\clearpage
\onecolumn

\begin{appendix}
\section{Catalog column description}
\label{Anexa1}

\begin{table*}[h]
\caption{Detections catalog: MOVIS-D.For the description of the columns 9-135 refer to the vhsDetection table description provided by VISTA Science Archive team on the site \url{$http://horus.roe.ac.uk/vsa/www/vsa_browser.html$}
}
\begin{tabular}{l l p{12cm}} \hline
Col. &Name           &Description \\  \hline
1 &Flag              & Each detection is flagged (first column) with the following code (corresponding to the flag variable):
\begin{itemize}
 \item 0 - indicates all valid measurements.These measurements passed all sorting criteria and were used for generating the MOVIS-C, and MOVIS-M catalog  
 \item 1 - indicates measurements for which the movement rate multiplied by the stack time interval ({\emph trail}) is above the specified threshold (1\arcsec) 
 \item 2 - indicates all the measurements which failed the {\emph O-C} criteria 
 \item 3 - indicates single night detections for the corresponding object. These measurements can not be used for color computation or for deriving photometric properties.
 \item 4 - indicates measurements for which the object has neighbors in the searched box. If the object is in only detected alongside neighbors in one of the stack frames from that night, then all the measurements from the set were indicated with code 4;
 \item 5 - indicates measurements for which the object overlaps with a background source, based on the comparison with USNO-B1 catalog
 \item 8 - indicates deprecated measurements, i.e. those measurements which are shown as deprecated in vhsDetection catalog or have saturation issues, or were shown as noise, or are signalized with post-processing error bits.  The flag is computed taking into account the following columns from the vhsDetection table: {\emph deprecated}, {\emph saturatCorr}, {\emph ppErrBits}, {\emph class}. For particular purposes, some of these measurements, which have low degree of deprecation, can be useful, but they were not included in our post-processing routines.
\end{itemize}
\\
2 &Number            & Asteroid number (e.g. 9495), or "-" if not available. The comet discoverers (e.g. P/La Sagra) are provided for comets\\
3 &Name              & Asteroid name (e.g. Eminescu) or provisional designation (2011 FP5) if the asteroid was not named. The name is provided for the comets (e.g. P/2011 A2, 279P) \\
4 &ClassObj          & Object classification, obtained from the SkyBoT service, based on the characteristics of their orbits\\
5 &Mv                & Computed visual apparent magnitude of the object at the moment of observation\\
6 &TrailLength\_asec & Trail length: the movement rate multiplied by the time interval in which the stack frame was obtained \\
7 &TotalTime\_min    & The total time interval in which the stack frame was obtained\\
8 &MeanMJD           & The mean modified Julian day of each detection  \\ \hline
\end{tabular}
\label{MOVISDDes}
\end{table*}

\begin{table*}[h]
\caption{Magnitudes catalog: MOVIS-M}
\begin{tabular}{l l p{12cm}} \hline
Col. & Name     & Description \\  \hline
1 &Number       & Asteroid number (e.g. 9495), or "-" if not available. The comet discoverers (e.g. P/La Sagra) are provided for comets\\
2 &Name         & Asteroid name (e.g. Eminescu), or provisional designation (2011 FP5) if the asteroid was not named. The name is provided for the comets (e.g. P/2011 A2, 279P) \\
3 &Nidx         & Index for marking the set of observations performed on different nights\\
4 &ClassObj     & Object classification, obtained from SkyBoT, based on the characteristics of the objects orbits\\
5 &Mv           & Computed visual apparent magnitude of the object at the moment of observation\\
6 &Phase        & The phase angle of the object at the moment of observation\\
7 &Nfilt        & The number of filters in which the object was observed (considering only the valid measurements)\\
8 &tinterv\_min & The time interval, given in minutes, in which the set of observations provided on each line were obtained \\
9 &meanMJD      & Mean modified Julian day of the set of observations provided on each line \\
10&YaperMag3    & Calibrated and corrected aperture magnitude in Y band. The 1\arcsec aperture radius was considered. If the observation was not available, the cell is left  empty \\
11&YaperMag3err & Error for the provided calibrated Y band aperture magnitude.  If the observation was not available, the cell is left  empty \\
12&YTrail       & The object movement rate multiplied by the the time interval in which the Y stack frame was obtained. The result is provided in arc seconds \\
13&YhlSMnRadAs  & Half-light semi-minor axis of the detection in Y band \\
14&JaperMag3    & Calibrated and corrected aperture magnitude in J band. The 1\arcsec radius aperture was used. If the observation was not available, the cell is left  empty \\
15&JaperMag3err & Error for the provided calibrated J band aperture magnitude.  If the observation was not available, the cell is left  empty \\
16&JTrail       & The object movement rate multiplied by the the time interval in which the J stack frame was obtained. The result is provided in arc seconds \\
17&JhlSMnRadAs  & Half-light semi-minor axis of the detection in J band \\
18&HaperMag3    & Calibrated and corrected aperture magnitude in H band. The 1\arcsec radius aperture was used. If the observation was not available, the cell is left  empty \\
19&HaperMag3err & Error for the provided calibrated H band aperture magnitude.  If the observation was not available, the cell is left  empty \\
20&HTrail       & The object movement rate multiplied by the the time interval in which the H stack frame was obtained. The result is provided in arc seconds \\
21&HhlSMnRadAs  & Half-light semi-minor axis of the detection in H band\\
22&KaperMag3    & Calibrated and corrected aperture magnitude in Ks band. The 1\arcsec radius aperture was used. If the observation was not available, the cell is left  empty\\
23&KaperMag3err & Error for the provided calibrated Ks band aperture magnitude.  If the observation was not available, the cell is left  empty \\
24&KTrail       & The object movement rate multiplied by the the time interval in which the Ks stack frame was obtained. The result is provided in arc seconds \\
25&KhlSMnRadAs  & Half-light semi-minor axis of the detection in Ks band\\ \hline
\end{tabular}
\label{MOVISFDes}
\end{table*}

\begin{table*}[h]
\caption{The colors catalog: MOVIS-C }
\begin{tabular}{l l p{12cm}} \hline
Col. & Name  & Description \\  \hline
1 &Number    & Asteroid number (e.g. 9495), or "-" if not available. The comet discoverers (e.g. P/La Sagra) are provided for comets\\
2 &Name      & Asteroid name (e.g. Eminescu), or the provisional designation (2011 FP5) if the asteroid was not named. The name is provided for the comets (e.g. P/2011 A2, 279P) \\
3 &ClassObj  & Object classification, obtained from SkyBoT. For asteroids, this classification is according to the characteristics of their orbits\\
4 &YmJ       & {\emph (Y-J)} magnitude\\
5 &YmJerr    & {\emph (Y-J)} magnitude error\\
6 &YmJt\_min & The time interval, given in minutes, between the observation in Y band and the observation in J band, used to compute Y-J\\
7 &YmH       & Y-H magnitude\\
8 &YmHerr    & Y-H magnitude error\\
9 &YmHt\_min & The time interval, given in minutes, between the observation in Y band and the observation in H band, used to compute Y-H\\
10&YmK       & Y-Ks magnitude\\
11&YmKerr    & Y-Ks magnitude error\\
12&YmKt\_min & The time interval, given in minutes, between the observation in Y band and the observation in Ks band, used to compute Y-Ks\\
13&JmH       & {\emph (J-H)} magnitude\\
14&JmHerr    & {\emph (J-H)} magnitude error\\
15&JmHt\_min & The time interval, given in minutes, between the observation in J band and the observation in H band, used to compute J-H\\
16&JmK       & J-Ks magnitude\\
17&JmKerr    & J-Ks magnitude error\\
18&JmKt\_min & The time interval, given in minutes, between the observation in J band and the observation in Ks band, used to compute J-Ks\\
19&HmK       & H-Ks magnitude\\
20&HmKerr    & H-Ks magnitude error\\
21&HmKt\_min & The time interval, given in minutes, between the observation in H band and the observation in Ks band, used to compute H-Ks\\
\hline
\end{tabular}
\label{MOVISCDes}
\end{table*}

{
\newpage
\clearpage
\section{Colors of Bus-DeMeo taxonomic templates spectra}
\label{Anexa2}

 \begin{table*}[h!] 
\caption{ Colors of Bus-DeMeo taxonomic templates spectra. The standard deviations were obtained from the computation of colors for the spectra used by \citep{2009Icar..202..160D} to define the taxonomy.}
\centering
\begin{tabular}{c c c c c c c c c c c c c}\hline
Class&$(Y-J)$&$\sigma_{Y-J}$&$(Y-H)$&$\sigma_{Y-H}$&$(Y-Ks)$&$\sigma_{Y-Ks}$&$(J-H)$&$\sigma_{J-H}$&$(J-Ks)$&$\sigma_{J-Ks}$&$(H-Ks)$&$\sigma_{H-Ks}$\\ \hline
A&0.461&0.038&1.226&0.146&1.359&0.161&0.765&0.127&0.898&0.152&0.133&0.051 \\
B&0.154&0.018&0.368&0.050&0.417&0.072&0.214&0.035&0.263&0.058&0.049&0.023 \\
C&0.209&0.020&0.529&0.044&0.701&0.068&0.320&0.025&0.493&0.050&0.173&0.028 \\
Cb&0.231&0.008&0.549&0.011&0.715&0.030&0.318&0.005&0.484&0.021&0.166&0.020 \\
Cg&0.226&  -  &0.545&  -  &0.687&  -  &0.320&  -  &0.461&  -  &0.142&  -   \\
Cgh&0.244&0.017&0.535&0.046&0.628&0.047&0.291&0.032&0.385&0.035&0.094&0.013 \\
Ch&0.233&0.022&0.525&0.044&0.626&0.061&0.292&0.026&0.393&0.048&0.101&0.026 \\
D&0.329&0.027&0.725&0.068&0.921&0.095&0.396&0.044&0.593&0.074&0.197&0.038 \\
K&0.269&0.026&0.601&0.065&0.690&0.079&0.332&0.043&0.421&0.059&0.089&0.024 \\
L&0.257&0.030&0.540&0.063&0.579&0.095&0.283&0.037&0.322&0.078&0.039&0.052 \\
O&0.583&  -  &0.980&  -  &0.870&  -  &0.397&  -  &0.287&  -  &-0.110&  -  \\ 
Q&0.380&0.038&0.764&0.046&0.826&0.076&0.385&0.030&0.446&0.053&0.062&0.033 \\
R&0.484&  -  &0.844&  -  &0.943&  -  &0.360&  -  &0.459&  -  &0.099&  -   \\
S&0.344&0.031&0.666&0.082&0.722&0.091&0.322&0.056&0.379&0.065&0.057&0.028 \\
Sa&0.358&0.049&0.956&0.053&1.111&0.008&0.598&0.004&0.753&0.057&0.156&0.061 \\
Sq&0.365&0.047&0.742&0.099&0.805&0.107&0.377&0.057&0.441&0.689&0.064&0.027 \\
Sr&0.381&0.039&0.693&0.080&0.740&0.096&0.311&0.048&0.359&0.075&0.048&0.039 \\
Sv&0.380&0.005&0.649&0.011&0.675&0.027&0.268&0.006&0.294&0.022&0.026&0.016 \\
T&0.261&0.005&0.594&0.008&0.750&0.011&0.333&0.008&0.489&0.013&0.157&0.009 \\
V&0.625&0.095&0.708&0.095&0.660&0.091&0.082&0.095&0.035&0.111&-0.047&0.050 \\
X&0.257&0.005&0.597&0.016&0.769&0.039&0.340&0.020&0.512&0.043&0.173&0.024 \\
Xc&0.228&0.012&0.513&0.025&0.624&0.057&0.285&0.013&0.396&0.047&0.111&0.034 \\
Xe&0.234&0.017&0.527&0.032&0.639&0.057&0.293&0.022&0.406&0.045&0.113&0.026 \\
Xk&0.257&0.019&0.564&0.050&0.692&0.086&0.307&0.034&0.435&0.071&0.129&0.039 \\

\hline
\end{tabular}
\label{TaxonColor}     
\end{table*}
}

\end{appendix}
\end{document}